%% file: STau3SC_NPJ_arxiv_03_08_2017.tex
\documentclass[aps,superscriptaddress,amsmath,amssymb,floatfix,twocolumn,prb, longbibliography]{revtex4-1}

\usepackage{amsmath}
\usepackage{amsfonts}
\usepackage{amssymb}
\usepackage{graphicx}

\usepackage{mathtools}
\usepackage{mathcomp}
\usepackage[caption=false,justification=raggedright,position=top,singlelinecheck=off]{subfig}
\usepackage{soul}
\usepackage[]{standalone}

\begin{document}
\title{Orbital selective pairing and superconductivity in iron selenides}

\author{Emilian M.\ Nica}
\affiliation{Department of Physics and Astronomy and Quantum Materials Institute,
University of British Columbia, Vancouver, B.C., V6T 1Z1, Canada}
\author{Rong Yu}
\affiliation{Department of Physics, Renmin University of China, 59 Zhongguancun St, Beijing, China, 100872}
\author{Qimiao Si}
\affiliation{Department of Physics and Astronomy, Rice University, 6100 Main St, Houston, TX, 77005, USA}

\begin{abstract}
An important challenge in condensed matter physics is understanding iron-based superconductors. 
Among these systems, the iron selenides hold the record for highest superconducting transition temperature and pose especially striking puzzles regarding the nature of superconductivity. The pairing state of the alkaline iron selenides appears to be of $d$-wave type based on the observation of a resonance mode in neutron scattering, while it seems to be of $s$-wave type from the nodeless gaps observed everywhere on the Fermi surface (FS). Here we propose an orbital-selective pairing state, dubbed $s \tau_{3}$, as a natural explanation of these disparate properties. 
The pairing function, containing a matrix $\tau_{3}$ in the basis of $3d$-electron orbitals, does not commute with the kinetic part of the Hamiltonian. This dictates the existence of  both intraband and interband pairing terms in the band basis. A spin resonance arises from a $d$-wave-type sign change in the intraband pairing component
whereas the quasiparticle excitation is fully gapped on the FS due to an $s$-wave-like form factor associated with the addition in quadrature of the intraband and interband pairing terms. We demonstrate that this pairing state is energetically favored when the electron correlation effects are orbitally selective. More generally, our results illustrate how the multiband nature of correlated electrons affords unusual types of superconducting states, thereby shedding new light not only on the iron-based materials but also on a broad range of other unconventional superconductors such as
heavy fermion and organic systems.
\end{abstract}

\date{\today}

\maketitle

\section{Introducton}

Unconventional superconductivity is driven by electron-electron interactions, 
instead of electron-phonon couplings~\cite{Anderson}. 
It occurs in a variety of strongly correlated electron systems, with the 
iron-based superconductors 
(FeSCs)
representing a prototype case
~\cite{Hosono, Johnston2011, Si2016,Wang_Lee2011,HosonoKuroki.2015,Hirschfeld2016}. 
The field of 
FeSC started with most of the efforts being directed toward 
 the iron pnictide class.
The normal state was found to be a bad metal,
with room-temperature resistivity reaching the Mott-Ioffe-Regel limit \cite{Johnston2011, Qazilbash:Nat_Phys_2009},
suggesting the importance of electron correlations \cite{Si_Abrahams:PRL_2008,Yin_Haule_Kotliar:Nat_Phys_2011}. 
More recently, the focus has been shifted to iron selenide systems. The reasons are 
manifold.
They
have 
the highest 
$T_c$~\cite{Xue.2012, Shen.2014}, 
they show 
even stronger electron correlations,
and, as we discuss here, their superconductivity is highly unusual.

The puzzle 
of the superconducting pairing state is highlighted 
by the ``122" alkaline iron selenides.
These systems have 
a
$T_c$
of about 
$31$ K at ambient pressure.
They have only 
electron Fermi pockets,
 lacking the hole pockets that exist in the iron pnictides 
at the center of the Brillouin Zone (BZ) \cite{Mou_et_al:PRL_2011,Zhang_et_al:PRL_2014,Wang_et_al:EPL_2011}.
Angle-resolved photoemission spectroscopy (ARPES)
 experiments 
 show that the quasiparticle dispersion is fully gapped 
on 
all the parts of the 
FS~\cite{Mou_et_al:PRL_2011,Zhang_et_al:PRL_2014,Wang_et_al:EPL_2011}, 
including a small electron Fermi 
pocket at the center of the BZ \cite{Xu_et_al:PRB_2012,Wang_et_al:EPL_2012}.
This is compatible with the usual $s$-wave $A_{1g}$ pairing state, 
but not with
the usual
 $d$-wave $B_{1g}$ state
 (which would produce nodes on the small electron Fermi pocket near the center of 
 the BZ).
On the other hand,
inelastic
neutron scattering experiments \cite{Park_etal:2011,Friemel_etal:2012}
observe a sharp resonance peak around the
wavevector $(\pi, \pi/2)$.
It is 
consistent 
with a pairing function that changes sign~\cite{Eschrig:Adv_Phys_2006}
 between the two Fermi pockets at the edge of the BZ, 
 such as would occur in a $d$-wave $B_{1g}$ state, but 
not in the usual $s$-wave $A_{1g}$ case.

In this work, we demonstrate how an orbital-selective pairing state, dubbed $s \tau_{3}$, 
exhibits properties that are commonly associated with a
$d$-wave $B_{1g}$ state 
\emph{or}
a $s$-wave $A_{1g}$ state. The key to the emergence of this superconducting state is the multiband nature
 of the FeSCs.
 This is associated with the multiplicity of $3d$ electron orbitals,
 whose conceptual importance
follows the tradition wherein new physics develops out of extra degrees of freedom, 
similar, for instance, to the way the so-called valley quantum number
in the electronic structure introduces new 
topological properties~\cite{Schaibley_2016}.
It is important for the FeSCs that there are multiple orbitals at play in the neighborhood of the Fermi level. 
Thus there is reason to expect that correlation effects  will be different for different orbitals.
In fact, there is evidence for orbitally-selective Mott behavior
 in the iron selenides ~\cite{MYi.2015, Yi_et_al:PRL_2013,Wang2014,Ding2014,Li2014}
and, thus, orbital selectivity is to be expected for pairing as well.

For strongly correlated superconductivity,
 Cooper pairing
 is naturally considered in an orbital basis 
  due to the 
  tendency of the electrons to avoid the dominating Coulomb repulsions.
Considering a basis
formed from all  
five 
$3d$-orbitals, the $s \tau_{3}$ 
state has an $s$-wave form factor, but 
transforms
as a $d$-wave $B_{1g}$ state. As such, it represents an energetically-favored reconstruction of the 
conventional $s$-wave and $d$-wave pairing states 
when they are quasi-degenerate, due to 
frustrated antiferromagnetic interactions \cite{Yu_Nat_Comm:2013}.
The pairing function incorporates
a matrix $\tau_3$ in the $3d_{xz}, 3d_{yz}$ subspace, which does not commute with the kinetic term of the Hamiltonian.
Consequently, in the band basis, it 
must also have 
a matrix structure,
which contains both
intraband and interband terms.
This allows 
the intraband pairing component 
to have 
a $d$-wave sign change,
while the addition in quadrature
of the intraband
and interband pairing terms is nonzero everywhere on the FS.
Thereby,
the spin excitations show
a $(\pi, \pi/2)$ resonance 
while
the quasiparticle excitations as measured by ARPES 
are fully gapped on the Fermi surface.

\section{Result}

%\section{Orbital selectivity in the normal state of iron selenides}
%\label{Sec_osm}

{\it Orbital selectivity in the normal state of iron selenides:}~~
In the normal state,
 ARPES has provided evidence not only for the existence of the orbital degree of freedom but also for strong orbital-selective correlation effects
 in the iron selenides.
 These materials include the 
 alkaline iron selenides, the Te-doped ``11'' iron selenides FeSe,
 and the monolayer FeSe on the SrTiO$_{\rm 3}$ substrate~\cite{MYi.2015, Yi_et_al:PRL_2013,Wang2014,Ding2014,Li2014}.
The effective quasiparticle mass normalized by its non-interacting counterpart, $m^*/m_{band}$
is on the order of $3-4$ for the $3d_{xz,yz}$ orbitals, but is as large as $20$ for the $3d_{xy}$ orbital~\cite{MYi.2015,Yi_et_al:PRL_2013,Liu2015}.
Such orbital selectivity has also
 been the subject of extensive recent theoretical studies
~\cite{Yu_Si:2011, Yu_Zhu_Si:PRL_2013, deMedici:PRL_2014}. 
All of these aspects make it
natural to study orbital dependent 
~\cite{Yu_Zhu_Si:2014,Yin:Nat_Phys_2014, Ong_Coleman_Schmalian:2014}
and related ~\cite{HaoHu:2014}
superconducting pairing.
We are thus motivated to 
address
 the hitherto unexplored question, {\it viz.} whether there exists an orbital-selective pairing state  which can
reconcile the seemingly contradictory properties observed in the iron-selenide superconductors.
We also 
examine
the stability of such a pairing state 
at the level of an effective Hamiltonian for studying superconductivity, 
in which we
incorporate
the
orbital-selectivity
in the short-range
exchange interactions
(see Supplementary Information (SI)).

%\section{Orbital-selective $s\tau_3$ pairing state: a simplified case}
%\label{Sec_st3_2orbital}

\begin{figure}[!htb]
\centering
\includegraphics[width=0.75\columnwidth]{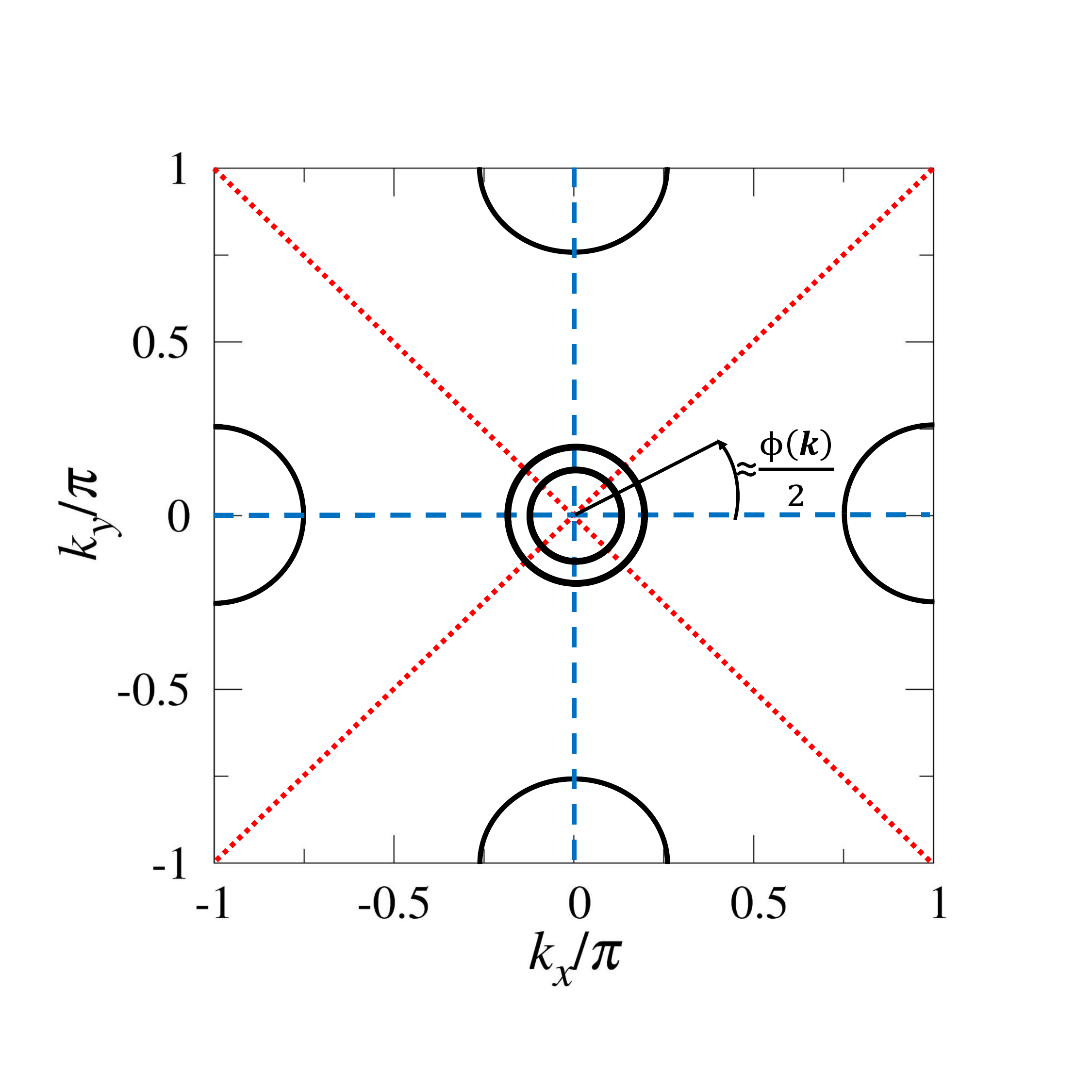}
%Empty_BZ_for_draft}
\caption{
\emph{Schematic} illustration of the two-orbital 
$s\tau_3$ pairing in a 1-Fe Brillouin Zone (BZ), which is 
obtained by unfolding the 2D crystallographic BZ cell in the conventional fashion~\cite{Nica_Yu_Si:Unpublished_2015}.
The solid lines indicate typical Fermi pockets for the Fe-based superconductors. 
The dotted, red lines indicate the zeroes specific to the intraband pairing (~$\xi_{-}$) while the dashed, blue lines mark the zeroes specific to the interband pairing (~$\xi_{xy}$).
The intra- and inter- band components do not vanish at the same subset of $\pmb{k}$, ensuring there is always a non-zero pairing given by either of the two components 
on the entire Fermi surface. 
For $\text{max}(\xi_{-}) \approx \text{max}(\xi_{xy})$ the angle $\phi(\pmb{k})$ 
(Eqs.~\ref{Eq:Balian_Wert_dispersion}-\ref{Eq:Ang_dep})
 can be roughly identified with twice the winding angle shown
 for fixed $|\pmb{k}|$. 
 In addition, there is a sign change between the intraband pairing along the two pockets at the edge of the BZ, a condition necessary to the formation of a resonance in the spin excitation spectrum at the wavevector $\pmb{q}=(\pi, \pi/2)$ observed in experiment \cite{Dai:arxiv_2015}.
}
\label{Fig:Empty_BZ}
\end{figure}

{\it Orbital-selective $s\tau_3$ pairing state -- a simplified case:}~~
We first discuss the structure and properties of the $s\tau_3$ pairing state in
a simplified two-orbital $d_{xz}, d_{yz}$ system. This
illustrates
how features typically associated 
with {\it both} standard structure-less $s$- and $d$-wave states can 
{\it simultaneously} arise.
The salient features of the two-orbital model are illustrated in Fig.~\ref{Fig:Empty_BZ}.

We consider
spin-singlet pairing in the orbital basis,
in the case of two orbitals 
$3d_{xz}$, $3d_{yz}$ 
~\cite{Raghu_et_al:2008}. 
The 
 Hamiltonian, incorporating 
  the 
$s\tau_{3}$
pairing
term,
 is given by

\noindent \begin{align}
\label{Eq;Hamilt_orbital}
\hat{H}=& \sum_{\pmb{k}} \psi^{\dagger}_{\pmb{k}} \big( \hat{H}_{\text{Kinetic}}(\pmb{k}) 
+ \hat{H}_{\text{Pair}}(\pmb{k}) \big) \psi_{\pmb{k}} \notag \\
\hat{H}_{\text{Kinetic}}= & \big( \xi_{+}(\pmb{k}) \otimes \tau_{0} + \xi_{-}(\pmb{k}) \otimes \tau_{3} 
+ \xi_{xy}(\pmb{k})  \otimes \tau_{1} \big) \otimes \sigma_{0} \otimes \gamma_{3} \notag \\
\hat{H}_{\text{Pair}}= &  \Delta_{0} g_{x^2y^2}(\pmb{k}) \otimes \tau_{3} \otimes \sigma_{0} \otimes \gamma_{1},
\end{align}

\noindent where $\psi^{\dagger}
_{\pmb{k}}
=(c^{\dagger}_{\pmb{k} i \sigma}, c_{-\pmb{k} j \sigma'} (i\sigma_{2})_{\sigma' \sigma}  )$ is equivalent to a Nambu spinor 
where $i,j$ are orbital indices 
(SI Section).
The 
$\tau_{i}, \sigma_{i}$, and $\gamma_{i}, (i=0, \ldots, 4)$
 2 x 2 Pauli matrices 
represent orbital iso-spin, spin, and Nambu indices, respectively.
The $\xi_{+}, \xi_{-}$, and $\xi_{xy}$ factors appearing 
in the kinetic  part 
belong to the $A_{1g}, B_{1g}$, and $B_{2g}$ irreducible 
representations
of the $D_{4h}$ point-group.
 Their exact forms, as well as 
the resulting electron bands 
are given in the SI.

\noindent 

The even-parity, spin-singlet candidate 
$s\tau_{3}$
pairing function 
with non-trivial orbital structure 
is included in the $\hat{H}_{\text{Pair}}$
term in Eq.~\ref{Eq;Hamilt_orbital}.
While $\Delta_{0}$ is a (generally) complex number,
we choose a real amplitude for convenience.
The form factor 
$g_{x^{2}y^{2}}(\pmb{k})$ is parity-even and belongs to
the $A_{1g}$ representation of the $D_{4h}$ point group.
In the absence of spin-orbit coupling, the rotational properties of the $s\tau_{3}$ pairing 
 are of $B_{1g}$ symmetry. The latter
 is entirely determined by the 
tensor product of the $g_{x^{2}y^{2}}(\pmb{k})$
(s-wave)
form factor and the $\tau_{3}$ orbital matrix.
To illustrate,
 under a $C_{4z}$ rotation, the form-factor is invariant, while the $\tau_{3}$ matrix 
transforms as a rank-two $B_{1g}$ tensor representation of the point-group, i.e. it changes sign. 
We note that the anti-symmetry under exchange is guaranteed by the spin-singlet nature, together with the even-parity of the form factor.
Since the spin-structure is not essential 
for the following arguments, we shall henceforth omit 
the explicit $\sigma_{0}$ matrix.

The non-trivial characteristics of this pairing 
are consequences of the commutator $\left[\hat{H}_{\text{Kinetic}}, \hat{H}_{\text{Pair}} \right] \neq 0$
for \emph{general} momentum $\pmb{k}$. 
We use the notation of Ref.~\citenum{Ong_Coleman_Schmalian:2014},
 and rewrite the Hamiltonian Eq.~\ref{Eq;Hamilt_orbital}
 as follows

\noindent \begin{align}
\label{Eq:Balian_Wert_form}
\hat{H}=& \sum_{\pmb{k}} \psi^{\dagger}({\pmb{k}}) \big[ \left( \xi_{+}(\pmb{k}) \tau_{0} +\vec{B}({\pmb{k}}) \cdot \vec{\tau} \right) \otimes \gamma_{3} \notag \\
& + \left( \vec{d}({\pmb{k}}) \cdot \vec{\tau} \right) \otimes  \gamma_{1} \big] \psi({\pmb{k}}),
\end{align}

\noindent where

\noindent \begin{align}
\vec{B}(\pmb{k})= & \left( \xi_{xy}(\pmb{k}), 0, \xi_{-}(\pmb{k}) \right)
\nonumber
 \\
\vec{d}(\pmb{k})= &  \left( 0, 0, \Delta_{0} g_{x^2y^2}(\pmb{k}) \right)
 \label{Eq:Pairing_form}.
\end{align}

\noindent This is formally similar to 
a Balian-Werthamer form
~\cite{Balian_PR_1963,Leggett1975,Sigrist_Ueda:1991}
(see SI for more details),
with the $\vec{B}(\pmb{k})$ factor 
being
 analogous to a $\pmb{k}$-dependent spin-orbit coupling.
To account for the non-commuting $\hat{H}_{\text{Kinetic}}$ and $\hat{H}_{\text{Pair}}$,
we write
 the square of the Hamiltonian matrix:

\noindent \begin{align}
\label{Eq:Squared_Hamilt}
\hat{H}^2= & \sum_{\pmb{k}} \left[ \xi_{+}(\pmb{k}) \tau_{0} + \left( \vec{B}_{\pmb{k}} \cdot \vec{\tau} \right) \right]^2 \otimes \gamma_{0} + \left| \vec{d}(\pmb{k}) \right|^2 \tau_{0} \otimes \gamma_{0} \notag \\
 & + 2i \left( \vec{B}(\pmb{k}) \times \vec{d}(\pmb{k}) \right) \cdot \vec{\tau} \otimes i \gamma_{2}.
\end{align}

\noindent where the well-known relation 
$\big( \vec{a} \cdot \vec{\tau} \big) \big( \vec{b} \cdot \vec{\tau} \big)= \vec{a} \cdot \vec{b}+i \big( \vec{a} \times \vec{b} \big) \cdot \vec{\tau}$
was used. The first two terms, 
proportional to the $\gamma_{0}$ Nambu matrix,
are the squares of the 
kinetic
Hamiltonian 
and of a pairing contribution with no essential structure in orbital space, given by 
$\left| \vec{d}(\pmb{k}) \right|^2$. 
The latter is an effective amplitude of
the pairing interactions and, as such, is proportional to the square of the s-wave like $g_{x^{2}y^{2}}$ form factor, as can be seen from Eq.~\ref{Eq:Pairing_form}. 
Together with the 
kinetic
part,
it  amounts to the usual (and sole) contribution to the 
Bogoliubov-de Gennes (BdG) quasiparticle spectrum, 
whenever $\left[\hat{H}_{\text{Kinetic}}, \hat{H}_{\text{Pair}} \right] = 0$
for \emph{all} $\pmb{k}$.
The last term 
in Eq.~\ref{Eq:Squared_Hamilt}
reflects the non-commuting $\hat{H}_{\text{Kinetic}}$ and $\hat{H}_{\text{Pair}}$.
Since the Nambu matrices $\gamma_{0}$ and $i \gamma_{2}$ commute, 
$\hat{H}^{2}$ in Eq. \ref{Eq:Squared_Hamilt} 
can be easily expressed in block diagonal form
 (SI).
The resulting Bogoliubov-de Gennes (BdG) bands are given by

\noindent
\begin{widetext}
\begin{equation}
\label{Eq:Balian_Wert_dispersion}
E_{\pm}(\pmb{k}) = \sqrt{ \left ( \sqrt{\xi^{2}_{+}(\pmb{k}) + \left| \vec{d}(\pmb{k}) \right|^2 \text{sin}^2\phi(\pmb{k}) } \pm \left| \vec{B}(\pmb{k}) \right|  \right)^2 + \left| \vec{d}(\pmb{k}) \right|^2 \left( 1- \text{sin}^2\phi(\pmb{k}) \right) }
\end{equation}

\noindent where

\noindent 
\begin{equation}
\label{Eq:Winding_angle}
\text{sin}\phi(\pmb{k})=\frac{\xi_{xy}(\pmb{k})}{\left| \vec{B}(\pmb{k}) \right|}= \frac{\xi_{xy}(\pmb{k})}{\sqrt{ \xi^2_{-}(\pmb{k})+ \xi^2_{xy}(\pmb{k})}}.
\end{equation}
\end{widetext}

\noindent 
The terms proportional to $\sin \phi(\pmb{k})$
reflects the non-Abelian aspect of the pairing state.
Note that
Eq.~\ref{Eq:Balian_Wert_dispersion} corresponds 
to the sum of two positive semi-definite terms. 
For \emph{general} $\vec{d}(\pmb{k})$ we see that
nodes can appear only when both terms in the square root vanish. 
The second of these goes to zero when either $\text{sin}\phi(\pmb{k})=1$ 
or, trivially, when $\left| \vec{d}(\pmb{k}) \right| = 0$. 
This latter case occurs when the FS intersects 
the lines of zeros of the $g_{x^{2}y^{2}}$ form factor. 
With the FeSCs in mind, we ignore this simple case in the following. Alternately,
when $\text{sin}\phi(\pmb{k})=1$, the dispersion reduces to

\noindent \begin{equation}
\label{Eq:Ang_dep}
E_{\pm}(\pmb{k}) = \left| \sqrt{\xi^{2}_{+}(\pmb{k}) + \left| \vec{d}(\pmb{k}) \right|^2 } \pm \left| \vec{B}(\pmb{k}) \right| \right|.
\end{equation}

\noindent \emph{On the FS}, we have 
$\xi^{2}_{+}(\pmb{k})=\left| \vec{B}(\pmb{k}) \right|^2$ 
(see SI).
Thus, 
there are no nodes \emph{on the FS}.

We note that \emph{away from the FS}, 
Eq.~\ref{Eq:Ang_dep} does not in general guarantee the absence of nodes. 
However, because the lifetime of quasiparticles away from the FS will be finite, 
the corresponding contributions to thermodynamical properties will be much weaker compared to the case of nodes on the FS. 

In the band basis, the kinetic part of the Hamiltonian is diagonalized. 
Given that the kinetic and pairing parts do not commute with each other,
the two cannot be simultaneously diagonalized.
Thus, the pairing part must contain an interband component.
To see this, 
we apply a 
canonical transformation which diagonalizes the kinetic part  (see the SI),
but which also transforms the pairing into
\noindent 
%\begin{widetext}
\begin{align}
\label{Eq:Pairing_band}
\hat{H}_{\text{Pair}}(\pmb{k})=\Delta_1(\pmb{k}) \alpha_3 + \Delta_2 (\pmb{k}) \alpha_1
\end{align}
%\end{widetext}
\noindent where $\alpha_{1,3}$ are 
Pauli matrices 
corresponding to inter- and intra-band pairing terms. The two components are given by

\noindent \begin{align}
\Delta_1(\pmb{k}) = &
-\Delta_0 g_{x^2y^2}(\pmb{k}) \frac{\xi_{-}(\pmb{k})}{\sqrt{\xi^2_{-}(\pmb{k})+\xi^2_{xy}(\pmb{k})}}
\nonumber
 \\
 \Delta_2(\pmb{k}) = &
 -\Delta_0 g_{x^2y^2}(\pmb{k})  \frac{\xi_{xy}(\pmb{k})}{\sqrt{\xi^2_{-}(\pmb{k})+\xi^2_{xy}(\pmb{k})}}
 \label{Eq:Delta-1-2}.
\end{align}

\noindent 

The band-diagonal $\alpha_3$ and band off-diagonal $\alpha_1$
pairing components have $d(x^2-y^2)$ and $d(xy)$ form factors, respectively.
As illustrated in 
Fig.~\ref{Fig:Empty_BZ}, 
these 
have nodes along 
the diagonals and axes of the BZ,
respectively.
Because the two matrices  $\alpha_{1,3}$ 
anti-commute, the single-particle excitation energy depends on the addition in quadrature of 
the two pairing amplitudes $\Delta_1(\pmb{k})$ and $\Delta_2(\pmb{k})$.
This ensures that the excitation gap is nodeless on the entire Fermi surface.

As can be seen from Eqs.~\ref{Eq:Pairing_band},~\ref{Eq:Delta-1-2}, 
the band-index diagonal term changes sign about the diagonals 
($k_x=\pm k_y$) of the BZ,
as dictated by the $d(x^2-y^2)$ nature of the intraband component.
Thus, the intraband pairing component does indeed change sign 
between the two electron Fermi pockets at the BZ boundaries.
It ensures that this type of pairing is conducive 
to the formation of a resonance with a wavevector that connects the two electron Fermi pockets.

We stress that the two main features of the $s\tau_{3}$ pairing, i.e.
the formation of a gap on the FS and the sign-change in the intraband component, cannot be reconciled by the more typical pairing candidates, 
which lack an orbital structure. In the context of our two-orbital model, the $s \otimes \tau_{0}$ and $d \otimes \tau_{0}$ candidate states, corresponding to the typical orbitally-trivial s and d-wave pairings, commute with $\hat{H}_{\text{Kinetic}}$. Consequently, they are associated with intraband pairing only. As such, neither of the two types can induce a nodeless gap \emph{and} account for the sign change required for the spin-resonance. 

%\section{Orbital-selective $s\tau_3$ pairing state: the case of iron selenides}
%\label{Sec_st3_5orbital}

{\it Orbital-selective $s\tau_3$ pairing state -- the case of iron selenides:}~~
Superconductivity in the alkaline iron selenides, like in the related case of the iron pnictides, involves all five Fe-$3d$ orbitals. 
Thus, it is important to consider 
the five-orbital case to address i) whether
 the $s\tau_{3}$ pairing state is energetically favored compared to the more conventional pairing states and ii) whether it captures 
  the essential properties of this pairing state as they pertain to the iron selenide
superconductors.

To 
study the stability of
the $s\tau_{3}$ pairing state, we start
from 
two
previously
discussed aspects 
of the FeSCs. 
We do so in terms of
a strong-coupling approach to superconductivity, in light of 
the strong correlation effects 
~\cite{Si_Abrahams:PRL_2008,Yin_Haule_Kotliar:Nat_Phys_2011,Fang08,Xu08,KSeo,Moreo.2009,WQChen.2009,DHLee.2013,Berg.2009,Lv,Bascones,deMedici:PRL_2014}
that are especially clear-cut for the iron selenides
~\cite{MYi.2015,Yi_et_al:PRL_2013,Liu2015}.
This approach is described in the SI, with superconductivity
driven by short-range interactions. 
The latter include the antiferromagnetic interactions between
the nearest-neighbor (NN, $J_{1}^{\alpha}$)
and next-nearest-neighbor (NNN,$J_{2}^{\alpha}$) Fe sites on their square lattice,
for the three 
most relevant orbitals,
$\alpha=$ $3d_{xz}$, $3d_{yz}$, and $3d_{xy}$.
We reiterate that we will analyze the model in the 1-Fe unit cell and the corresponding BZ.

One 
of the known
aspects of the FeSCs is the large parameter regime 
where 
 the conventional 
$d$-wave $B_{1g}$ and $s$-wave $A_{1g}$
pairing states 
are quasi-degenerate
~\cite{Yu_Nat_Comm:2013,Graser.2009}.
In terms of a model 
with short-range antiferromagnetic interactions, this occurs in the regime of magnetic 
frustration with $J_{2}$ being comparable to $J_{1}$~\cite{Yu_Nat_Comm:2013}, a condition that is evidenced by both
theoretical considerations and experimental measurements
~\cite{Si2016,Dai:arxiv_2015}.
 To quantify this effect, we introduce the ratio
 $A_{L} \equiv
 J_2/J_1$ to describe the relative strength of these two interactions.
 For a proof-of-concept demonstration, we analyze the phase diagram by taking the $J_2/J_1$ axis to be a cut in the parameter 
 space along which $A_L$ is the same for the different $3d$ orbitals.
 The quasi-degeneracy arises 
 when $A_{L} \sim 1$.
 
 The second well-known property of the FeSCs
is orbital selectivity, as 
described
above.
Our effective model incorporates
an exchange orbital-anisotropy factor
 $A_{O}=J_{1}^{xy}/J_{1}^{xz/yz}=J_{2}^{ xy}/J_{2}^{xz/yz}$,
 and reflects the orbital selectivity 
 by $A_{O}$'s deviation from $1$.
For the iron selenides, $A_O$ is expected to be considerably smaller than $1$ (see SI).

We are now in position to discuss how the
$s\tau_{3}$
pairing state emerges in a range of parameters where the $s-$ and $d-$wave pairing channels are quasi-degenerate.
Within the 5-orbital $t-J_{1}-J_{2}$ model, we focus on the case with a kinetic part appropriate for the alkaline iron selenides
K$_{y}$Fe$_{2-x}$Se$_{2}$ although similar behavior emerges in the cases appropriate for the iron pnictides and single-layer FeSe
(see SI).
We present our results for the case of orbital-diagonal exchange interactions. The inter-orbital exchange interactions have only negligible
effects on the pairing amplitudes, as demonstrated in the SI.

\begin{figure}[htb!]
\centering
\subfloat[]{\includegraphics[width=0.8\columnwidth]{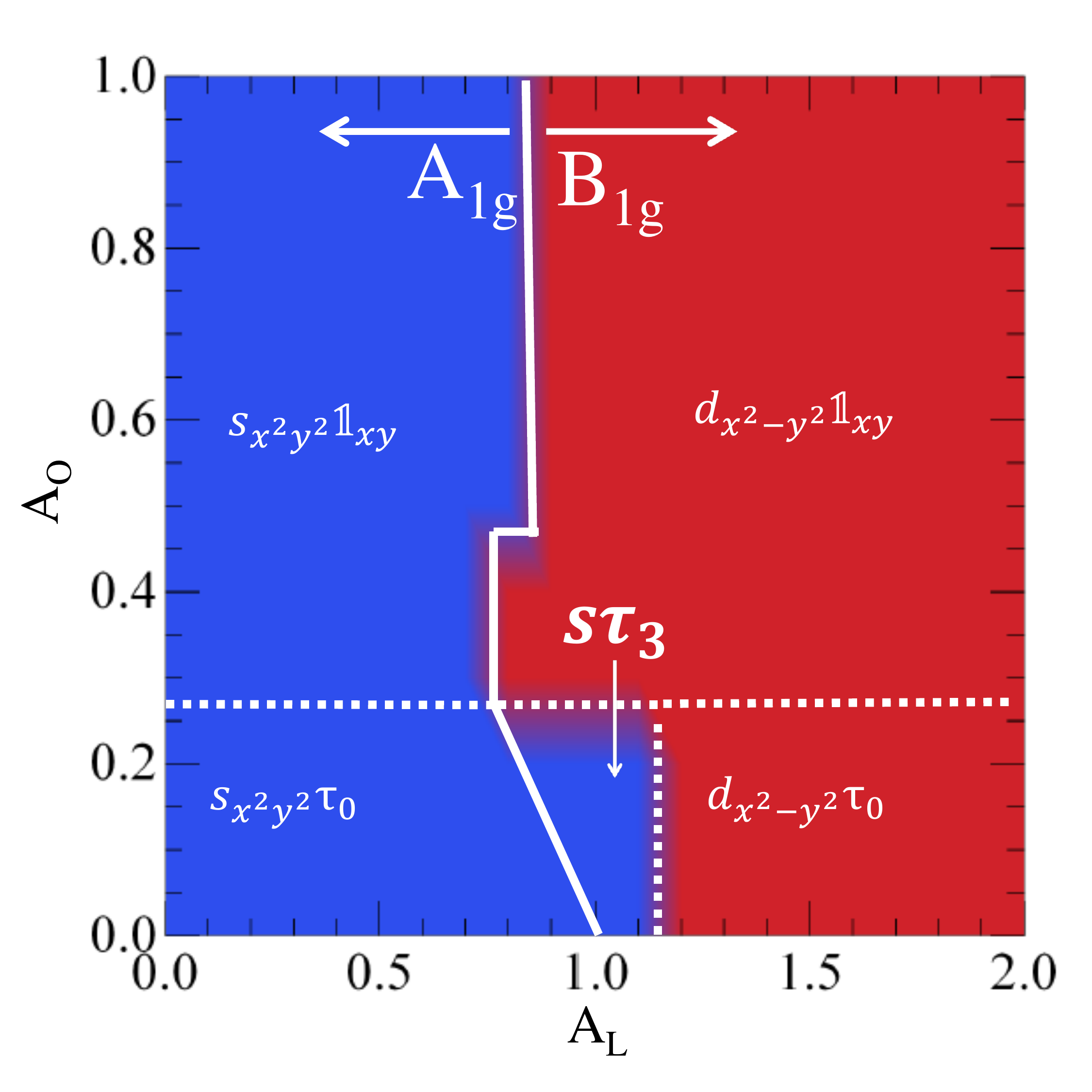}}\\
%Phase_diagram_K__PNAS_11_10_2016}}\\ 
\subfloat[]{\includegraphics[width=0.8\columnwidth]{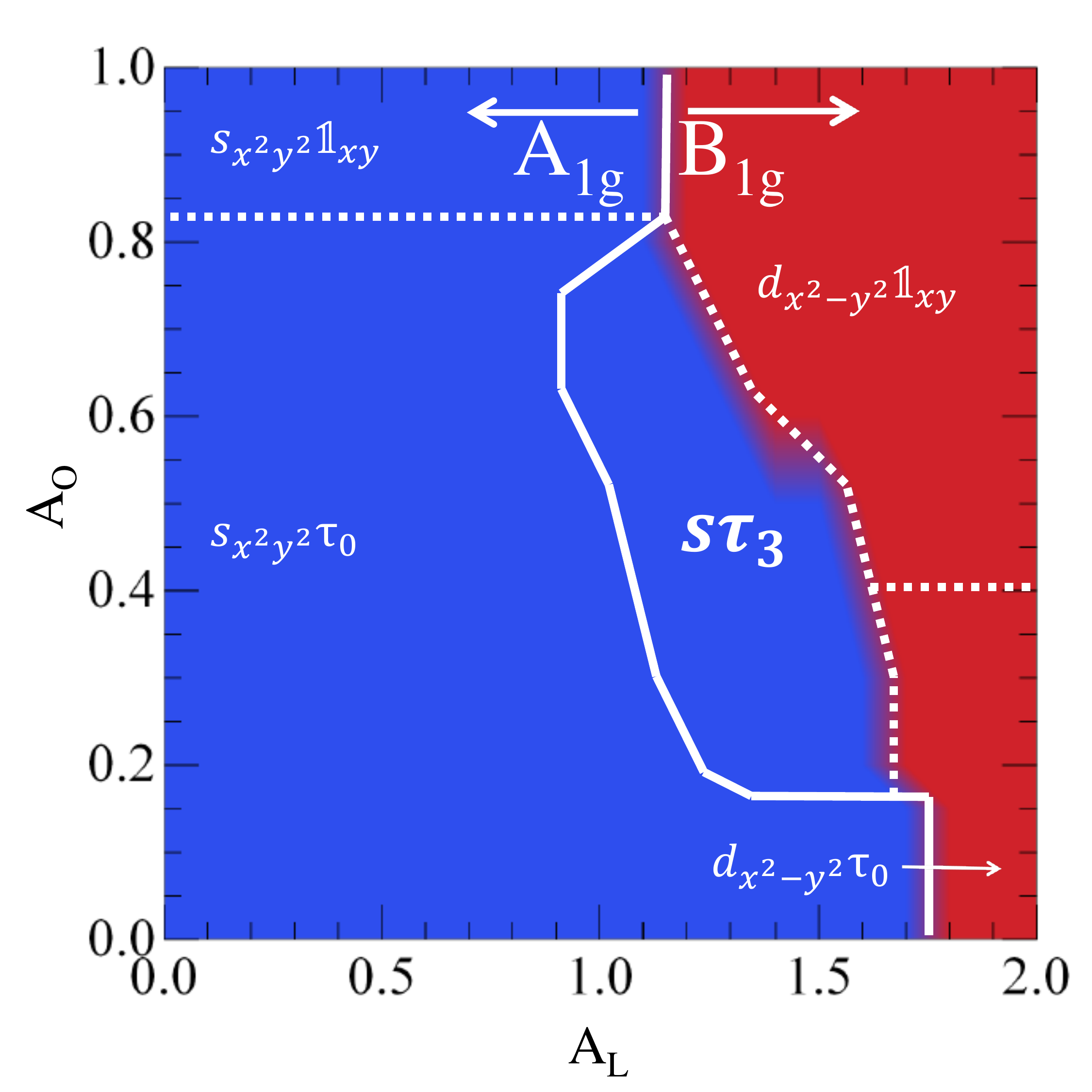}}
%Phase_diagram_P_PNAS_11_10_2016}}
\caption{
Phase diagrams based on the leading pairing amplitudes given by self-consistent calculations with fixed $J_2=1$ and tight-binding parameters appropriate to (a) alkaline iron selenides, and (b) iron pnictides. The tight-binding parameters used can be found in Ref. \citenum{Yu_Nat_Comm:2013}. The blue shaded areas correspond to dominant pairing channels with an $s_{x^2y^2}$ form factor while the red shading covers those with a $d_{x^2-y^2}$ form factor. The continuous line separates regions where the pairing belongs to the $A_{1g}$ and the $B_{1g}$ representations respectively. The $1 \times 1$ matrix in the $d_{xy}$ subspace is represented by $\pmb{1}_{xy}$. The orbital-selective $s\tau_{3}$ pairing occurs for $A_O < 1$, $A_L$ near 1 in all cases.
}
\label{Fig:Phase_diagrams_K_Fe_Se}
\end{figure}

The phase diagram for the alkaline iron selenides is shown in Fig.~\ref{Fig:Phase_diagrams_K_Fe_Se} (a).  
In the absence of orbital selectivity, $A_O=1$, it is known that 
small and large $A_L$ 
promote the $s_{x^2y^2} \otimes \tau_{0} , {A_{1g}}$ 
and $d_{x^2-y^2} \otimes \tau_{0}, B_{1g}$,
both defined in the $d_{xz}$, $d_{yz}$ subspace~\cite{Yu_Nat_Comm:2013}.
Increasing the orbital selectivity, with $A_O$ decreasing from $1$,
these two limiting regimes
remain essentially unchanged. 
However, in the magnetically frustrated regime
$A_L \sim 1$,
the $s_{x^2y^2} \otimes \tau_{0} , {A_{1g}}$ 
and $d_{x^2-y^2} \otimes \tau_{0}, B_{1g}$
become quasi-degenerate.
When $A_O$ is sufficiently smaller than $1$, the $s\tau_{3}$ pairing state becomes the dominant channel
in the intermediate regime.
Similar phase diagrams are obtained for the iron pnictides and single-layer FeSe  
shown in Figs.~\ref{Fig:Phase_diagrams_K_Fe_Se} (b) and~S1
(SI), respectively. 
A typical dominant
$s\tau_{3}$ pairing case is shown in 
Fig.~S2  in the 
SI
 for a number of subleading symmetry-allowed channels~\cite{Goswami_Nikolic_Si:EPL_2010} for alkaline iron selenide dispersion with fixed $J_2/J_1=1.5$, $A_O=0.3$ and varying $A_L$ (horizontal axis).

\begin{figure}[!htb]
\centering
\includegraphics[width=0.7\columnwidth]{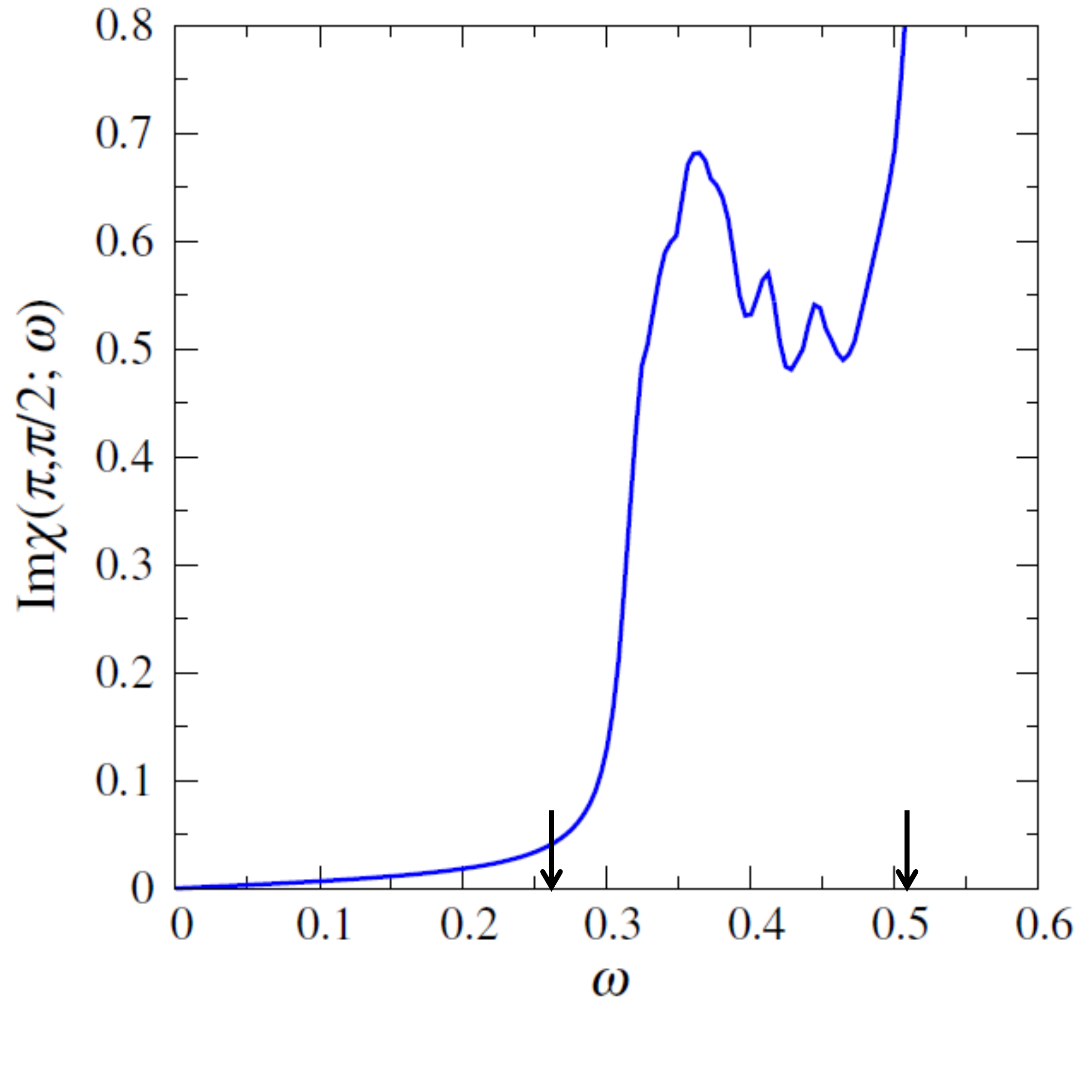}
%K_Im_Chi_pi_0_5_pi_J_2=1_5_A_L=0_9}
\caption{
The imaginary part of the dynamical spin susceptibility for the alkaline iron selenides at wave-vector $\pmb{q}=(\pi, \pi/2)$, for a dominant $s\tau_{3}$ pairing for parameters $J_2=1.5$, $A_O=0.3, A_L=0.9$. 
The arrows show twice the minimum and maximum gaps (see Fig. \ref{Fig:K_FS_pair_sign} (b)). There is a sharp feature ar $\omega \approx 0.36$ within the bounds of twice the effective gap and below the p-h threshold of roughly 0.41 associated with this wavevector.
}
\label{Fig:K_spin_spectrum}
\end{figure}

\begin{figure*}[htb!]
\centering
\subfloat[]{\includegraphics[width=0.75\columnwidth]{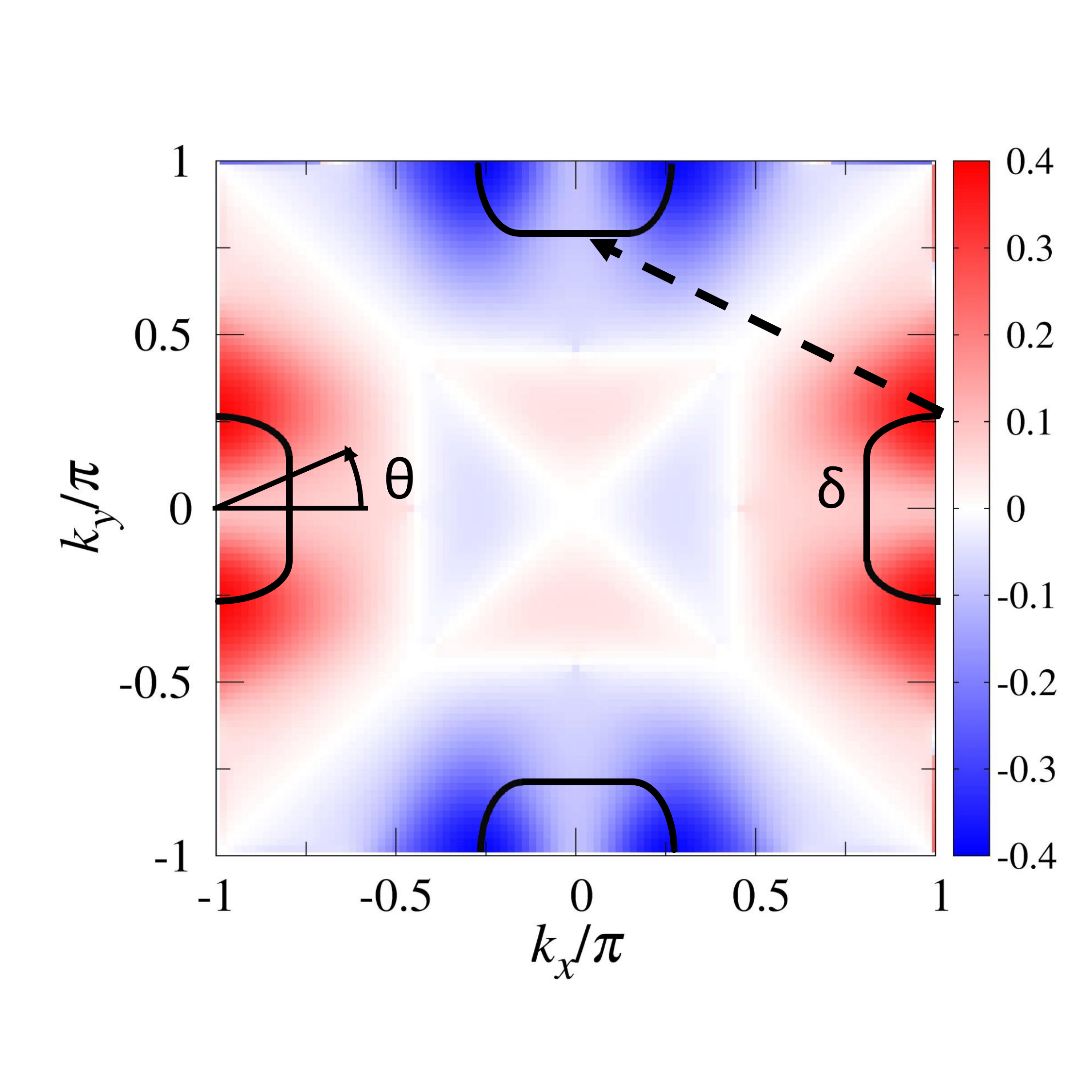}}
%K_real_pairing_band_44_J_2=1_5_A_L=0_9}} 
\qquad
\subfloat[] {\includegraphics[width=0.75\columnwidth]{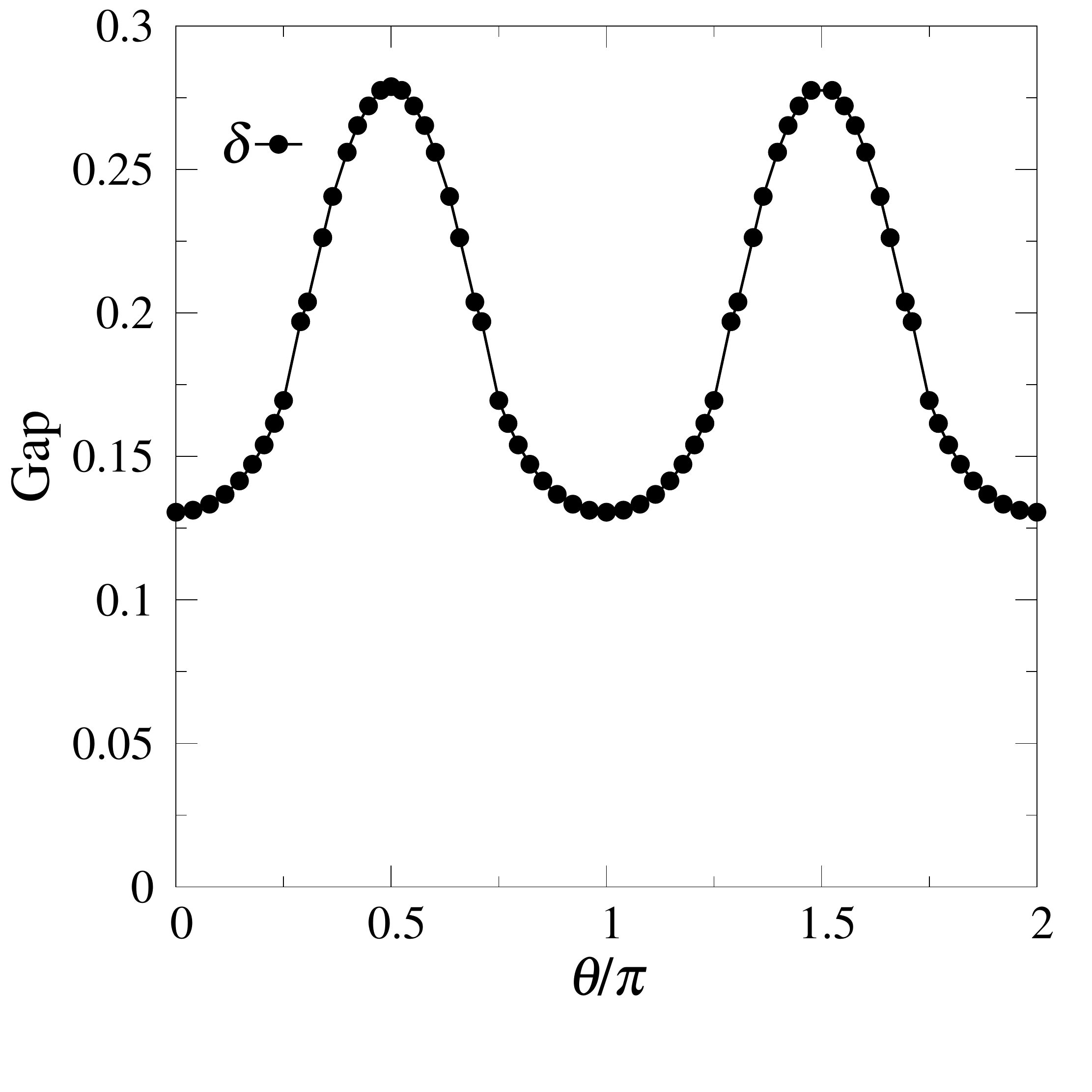}}
%KGapFSn615Ja090_J_2_1_one_half}}
\caption{
(a) The FS (solid line) and the real intra-band pairing for the band generating the $\delta$ pockets at the edge of the BZ for a dispersion typical of the alkaline iron selenides. Note the clear change in sign between pockets separated by the BZ diagonal. The dashed arrow indicates the $\pmb{q}=(\pi,\pi/2)$ wave-vector associated with the resonance in the spin spectrum found in experiment \cite{Dai:arxiv_2015}. (b) The size of the gap along the $\delta$ pocket. Both figures are for $J_2=1.5$, $A_O=0.3, A_L=0.9$ with dominant $s\tau_{3}$ pairing. 
}
\label{Fig:K_FS_pair_sign}
\end{figure*}

Having established the stability of the $s\tau_3$ pairing state, we now address its salient properties. We first consider the spin-excitation spectrum.
In Fig.~\ref{Fig:K_spin_spectrum} we show the 
dynamical spin susceptibility 
at wave-vector $\pmb{q}=(\pi, \pi/2)$ for $J_2=1.5$. We note the complicated frequency behavior which can be traced to the anisotropy in the effective gap affecting both the coherence factors and the position of minimum in quasi-particle energy. We show the minimum and maximum particle-hole (p-h)
 thresholds corresponding to twice the minimum and twice the maximum gaps. As suggested by Figs.~\ref{Fig:K_FS_pair_sign} (a) and (b), states connected by $\pmb{q}=(\pi, \pi/2)$ would correspond to a p-h threshold given roughly by the sum of the minimum and maximum gap.
 A sharp feature appears below this threshold,
 confirming the existence of the resonance for $\pmb{q}=(\pi, \pi/2)$ as found in experiments on the alkaline iron selenides \cite{Park_etal:2011,Friemel_etal:2012,Dai:arxiv_2015}.
The resonance at this wavevector originates from the sign change of the intraband pairing component across 
the two Fermi pockets at the edge of the BZ, around $(\pm \pi,0)$ ($\delta$) and $(0, \pm \pi)$, as illustrated in Fig.~\ref{Fig:K_FS_pair_sign} (a),
and further discussed in the SI.
Without such a sign change, there cannot be a sharp resonance 
{\it below} the p-h threshold energy.

\noindent 

We next turn to the quasiparticle excitation spectrum.
Fig.~\ref{Fig:K_FS_pair_sign} (b) shows the gap at the FS as a function of winding angle $\theta$. 
It clearly illustrates the node-less dispersion as the gap is nonzero for all $\theta$.

\begin{figure*}[htb!]
\centering
\subfloat[]{\includegraphics[width=0.75\columnwidth]{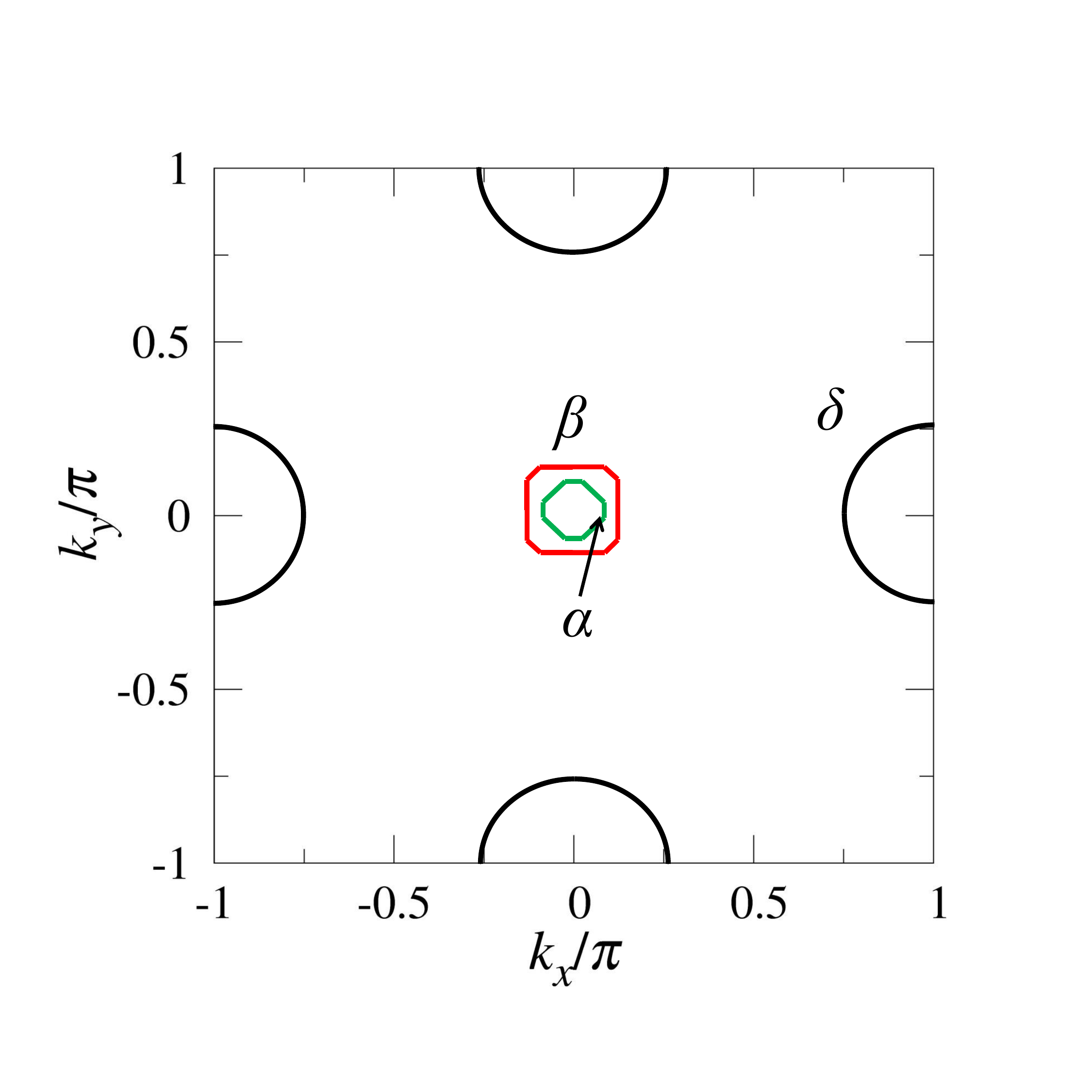}}
%AFSn615_resized}}
 \qquad
\subfloat[]{\includegraphics[width=0.75\columnwidth]{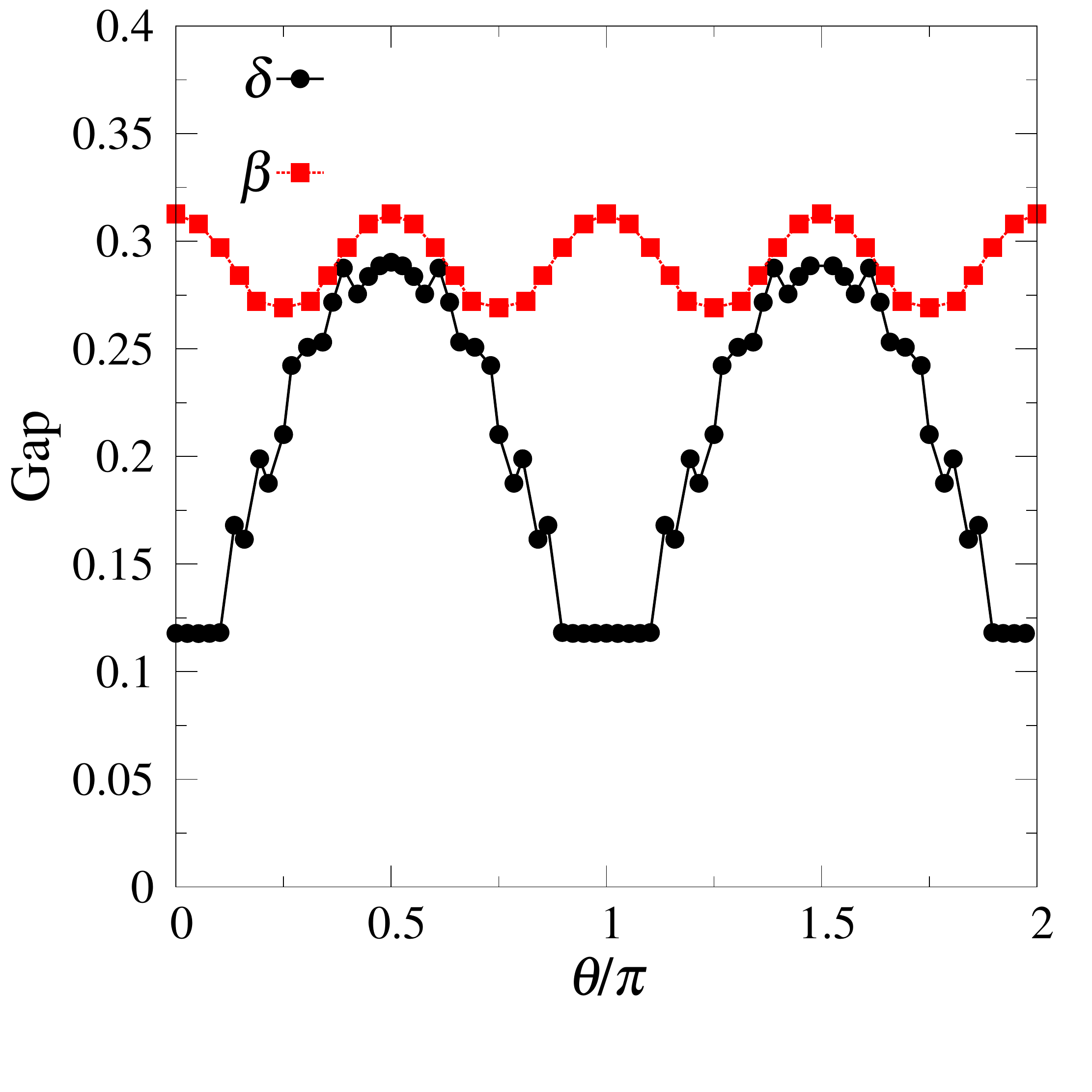}}
%AGapFSn615Ja130_resized}}
\caption{
(a) FS for the iron pnictides which includes hole pockets around the $\Gamma$ point
with dominant $s\tau_{3}, B_{1g}$  
for $J_2=1, A_L=1.3, A_O=0.5 $.The tight-binding parameterscan be found in Ref.~\citenum{Yu_Nat_Comm:2013}. (b) The gaps along the $\beta$ and $\delta$ pockets close to the center and edge of the BZ. A similar gap forms around the $\alpha$ pocket.
}
\label{Fig:A_FS_gap}
\end{figure*}

 The electron dispersion considered here does not produce any Fermi pockets close to $\Gamma$ in the BZ. This is in contrast to ARPES experiments on K$_{y}$Fe$_{2-x}$Se$_{2}$~\cite{Zhang_et_al:2011,Liu-et-al:2012} 
 which show a small electron pocket near $\Gamma$. Because this 
 electron pocket has very small spectral weight, it is to be expected that even if such a pocket were included, the dominant 
 $s\tau_{3}$ pairing will still arise; moreover, the gap on this Fermi pocket will be node-less as discussed in the two-orbital case. To substantiate this, 
 we consider the results for the iron pnictides class, which do have significant ({\it albeit} hole) Fermi pockets
 at the zone center yet exhibit a full gap. In Figs.~\ref{Fig:A_FS_gap} (a), (b) we show the FS and the gaps as functions of winding angle $\theta$ for $A_O=0.5$ and $A_L=1.3$ corresponding to a dominant
 $s\tau_{3}$ pairing. The gap along $\beta$ is finite and exhibits an anisotropy consistent with the two orbital results in 
 Eq.~\ref{Eq:Balian_Wert_dispersion}. In the latter case, at winding angle $\theta=0$, $\text{sin}\phi=0$ and the spectrum has a minimum/maximum gap for $E_{+/-}$. As $\theta$ is increased the $\left| \vec{B}(\pmb{k}) \times \vec{d}(\pmb{k}) \right|^2$ term increases reaching a maximum at $\theta=\pi/4$. Here the gap is maximum/minimum for $E_{+/-}$. This is consistent with the anisotropy in the gap shown in Fig.~\ref{Fig:A_FS_gap}. 

\noindent 

\noindent 

\section{Discussion}
\label{Sec:Dscssn}

Several remarks are in order. 
First, the full gap and the sign change of the intraband pairing component discussed above provide evidence that, with strong orbital selectivity, the $s \tau_{3}$ pairing in a realistic five-orbital 
model
has a behavior very similar to that of the two-orbital case. 

Second, with the short-range $J_1-J_2$ interactions driving superconductivity, 
pairing involves the electronic states over an extended range of energy
about the Fermi energy. 
The energy window can be determined from the zone-boundary spin excitation energies, 
which 
are 
on the order of $200$ meV for most iron selenides (and pnictides)~\cite{Dai:arxiv_2015}.
This is important for the consideration of the quasiparticle excitation gap at the small
electron pocket 
of K$_{y}$Fe$_{2-x}$Se$_{2}$
near the origin of the Brillouin zone.
According to the ARPES experiments~\cite{Zhang_et_al:2011, Liu-et-al:2012}
this Fermi pocket 
contains Fe $3d_{xy}$ and Se $4p_z$ orbitals ($\alpha$ band),
while
the hole ($\beta$) bands containing both $3d_{xz}$ and $3d_{yz}$ orbitals and are
only about $60$-$80$ meV below the Fermi energy.
We therefore expect that both the intraband and interband pairing components will be significant 
for this part of the Brillouin zone and the mechanism advanced here will make the quasiparticle excitations to be fully gapped for this small electron pocket.

Third, within our approach both the iron selenides and 
pnictides are bad metals in the regime of quasi-degenerate $s-$ and $d-$wave pairings. However, the iron selenides have stronger correlations, 
which will lead to a larger ratio of the exchange interaction to renormalized kinetic energy 
(note that the {\it renormalized} bandwidth goes to zero when a bad metal approaches the electron localization transition) and,
correspondingly~\cite{Yu_Nat_Comm:2013}, 
larger pairing amplitudes. We expect 
this will contribute to the larger maximum $T_c$ observed in the iron selenides than in the iron pnictides. 
Relatedly,
the alkaline iron selenides have a stronger orbital selectivity than the iron pnictides, 
and we thus expect that the $s\tau_3$ pairing 
is more likely 
 realized in the former than in the latter.

Fourth, it is instructive to 
compare the mechanism advanced here with 
a conventional means of relieving quasi-degenerate $s-$ and $d-$wave pairing states with trivial orbital structure, which
 consists in linearly superposing the two into an $s+id$ state.
 The latter, breaking 
the time-reversal symmetry, would be stabilized at temperatures sufficiently below the superconducting transition temperature. 
By contrast, the $s\tau_{3}$ pairing state 
preserves the time-reversal symmetry. It is an 
irreducible representation of the point group, and 
is therefore
stabilized as the temperature is lowered immediately below the superconducting transition.
Thus, the emergence of the intermediate $s\tau_{3}$ pairing state 
represents a new means to relieve the quasi-degeneracy through the development of orbital selectivity. 

Finally, the nodeless $d$-wave nature of $s\tau_{3}$ may shed new light on other strongly correlated multi-band superconductors. For instance, one of the striking puzzles emerging in heavy fermion superconductors is the simultaneous exhibition of a variety of $d-$wave characteristics and of a gap in the lowest-energy excitation spectrum~\cite{Kittaka-2014}. Whether a multiband pairing state such as $s\tau_{3}$ provides a systematic understanding of such properties is an intriguing open question for future studies.

%\section{Conclusion}
%\label{Conclusion}

%We
To summarize,
we have demonstrated that an orbital-selective $s\tau_{3}$ pairing state exhibits properties 
that would appear mutually exclusive from the conventional perspective where the orbital degrees of freedom are ignored. 
It provides a natural understanding of the enigmatic properties observed in the alkaline iron selenides. 
These include the single-particle excitations which are fully gapped on the entire Fermi surface, 
as observed in ARPES experiments, and a pairing function which changes sign across the electron Fermi 
pockets at the Brillouin-zone 
boundary, as indicated by the resonance peak seen near $(\pi,\pi/2)$ in the inelastic neutron scattering experiments. 
In addition, we have shown that the pairing state is energetically competitive in an orbital-selective model of short-range 
antiferromagnetic exchange interactions, in the regime where the conventional $s-$ and $d-$wave pairing channels are quasi-degenerate. 
As such, our understanding of the properties of the iron-selenide superconductors provides evidence 
that the high-T$_{\rm c}$ superconductivity in the iron-based materials originates from the antiferromagnetic correlations 
of strongly correlated electrons. More generally, our work highlights how new classes of unconventional superconducting pairing state 
emerge in the presence of additional internal degrees of freedom, with properties that cannot otherwise be expected. 
This new insight may well be important for the understanding of a variety of other strongly correlated superconductors, 
including the heavy fermion and organic systems.

%\showmatmethods % Display the Materials and Methods section

\section{Acknowledgments}
\label{Acknow}

We acknowledge useful discussions with E. Abrahams, A. V. Chubukov, G. Kotliar and P. J. Hirschfeld. 
The authors declare that they have no competing interests.
All authors contributed to the research of the work and the writing of the paper. 
The work has been supported in part by the NSF Grant No.\ DMR-1611392 and the Robert A.\ Welch Foundation Grant No.\ C-1411 
(E.M.N. \& Q.S.). R.Y. was partially supported by the National Science Foundation of China Grant number 11374361, 
and the Fundamental Research Funds for the Central Universities and the Research Funds of Renmin University of China. 
All of us acknowledge the support provided in part by the NSF Grant  No. NSF PHY11-25915 at KITP, UCSB.
Correspondence and requests for materials should be addressed to E.M.N. (enica@qmi.ubc.ca)
or Q.S. (qmsi@rice.edu).

\bibliography{tau_3_03_08_2017_revtex}

\widetext
\pagebreak
\begin{center}
\textbf{\large Orbital selective pairing and superconductivity in iron selenides:\\[+2ex] Supporting Information} \\[+2ex]
\text{Emilian M. Nica, Rong Yu, and Qimiao Si}
\end{center}

\setcounter{section}{0}
\setcounter{equation}{0}
\setcounter{figure}{0}
\setcounter{table}{0}
\setcounter{page}{1}
\makeatletter
\renewcommand{\theequation}{S\arabic{equation}}
\renewcommand{\thefigure}{S\arabic{figure}}
\renewcommand{\bibnumfmt}[1]{[S#1]}
\renewcommand{\citenumfont}[1]{S#1}

\input{STau3SC_SI_arxiv_03_08_2017_revtex}

\end{document}

%% file: STau3SC_SI_arxiv_03_08_2017_revtex.tex
\title{Orbital selective pairing and superconductivity in iron selenides - Supporting Information}
\author{Emilian M. Nica, Rong Yu, and Qimiao Si}

%\doi{10.1073/pnas.XXXXXXXXXX}

% Set the LAST citation number from your
% manuscript here
%\continuecitefrom{55}

% For references that are already cited in the
% main manuscript, need to override its number
% manually here, based on its number in the main
% paper. Then cite this reference in the SI with
% \cite.
%\defcitealias{Qazilbash:Nat_Phys_2009}{8}
%\defcitealias{Yu_Nat_Comm:2013}{27}
%\defcitealias{Sigrist_Ueda:1991}{41}
%\defcitealias{Yu_Zhu_Si:PRL_2013}{30}
%\defcitealias{Yu_Zhu_Si:2014}{32}
%\defcitealias{Raghu_et_al:2008}{36}
%\defcitealias{Nica_Yu_Si:Unpublished_2015}{37}
%\defcitealias{Eschrig:Adv_Phys_2006}{20}
%\defcitealias{Fong_et_al:PRL_1995}{56}

%\begin{document}

\maketitle

\section{Two-orbital model}
\label{Sec:Dscssn_tw_orbtl_mdl}

\subsection{Tight-binding details}

The components of the tight-binding part
of the two-orbital Hamiltonian discussed in the main text are given by

\noindent \begin{align}
\xi_{\pmb{k}+} = & -(t_1+t_2)(\cos k_x+\cos k_y) \notag \\
& -4t_3\cos k_x \cos k_y, \\
\xi_{\pmb{k}-}= & -(t_1-t_2)(\cos k_x-\cos k_y), \\
\xi_{\pmb{k}xy}= & -4t_4\sin k_x \sin k_y,
\end{align}

\noindent where $t_1$,$t_2$ and $t_3$ are tight-binding parameters.
Details can be found in Ref.~\onlinecite{S_Raghu_et_al:2008}.
The corresponding band dispersion is in general given by

\noindent \begin{align}
\label{Eq:Free-band}
\epsilon_{\pm}(\pmb{k})= & \xi_{+}(\pmb{k}) \pm \sqrt{\xi^2_{-}(\pmb{k})+ \xi^2_{xy}(\pmb{k})} \notag \\
= & \xi_{+}(\pmb{k}) \pm \left| \vec{B}(\pmb{k})\right|
\end{align}

\noindent The Fermi surface is determined by the condition

\noindent \begin{equation}
\epsilon_{\pm} (\pmb{k}_{FS}) = 0,
\end{equation}

\noindent which is equivalent to

\noindent \begin{equation}
\xi_{+}(\pmb{k}_{FS}) = \mp \left| \vec{B}(\pmb{k}_{FS}) \right|.
\end{equation}

\subsection{Nambu form}

The pairing part
written as $\hat{H}_{\text{Pair}} \sim \vec{d} \cdot \vec{\tau}$
 is equivalent to a more-conventional Balian-Werthamer form
$ \left( \tilde{\pmb{d}} \cdot \vec{\tau} \right) \left(i \tau_{2} \right)$
which is conventionally used for pairing functions with non-trivial spin structure.
This is so provided that $d_{2}=\tilde{d}_{2}=0$, which is the case for $s \otimes \tau_{3}$ pairing,
together with ${d}_{1} \rightarrow \tilde{d}_{3}, {d}_{3}\rightarrow  -\tilde{d}_{1}$.
Formally, this transforms $2i \left( \vec{B} \times \vec{d}\right) \cdot \vec{\tau}$
in the expression for $\hat{H}^{2}$ (Eq.~4 in the main text)
%\ref{Eq:Squared_Hamilt})
 to
$ 2 \left( \vec{B} \cdot \pmb{\tilde{d}} \right) i\tau_{2}$. The resulting BdG bands are
identical, as can be seen by expanding the direct products.
Note that, in contrast to the typical \emph{spin}-triplet pairing,
both $\vec{d}$ and $\tilde{d}$ orbital iso-spin vectors
are parity-even $\left(\vec{d}(-\pmb{k})=\vec{d}(\pmb{k})\right)$.
Together with the spin-singlet nature,
this ensures that the Cooper pairs are anti-symmetric under exchange.

In order to better illustrate the effects of the non-trivial orbital structure,
we incorporate the spin-singlet nature of the pairing Hamiltonian into a transformed Nambu spinor:

\noindent \begin{align}
\psi^{\dagger}(\pmb{k})= \psi^{\dagger}_{\text{Nambu}}~U^{\dagger}
\end{align}

\noindent where

\noindent \begin{equation}
\psi^{\dagger}_{\text{Nambu}}(\pmb{k})=(c^{\dagger}_{\pmb{k} i \sigma}, c_{-\pmb{k} j \sigma}),
\end{equation}

\noindent is the canonical Nambu spinor and

\noindent \begin{equation}
U=\frac{1}{2} \big( \sigma_{0} \otimes (\gamma_{0}+\gamma_{3})
+  i\sigma_{2} \otimes (\gamma_{0}-\gamma_{3}) \big) \otimes \tau_{0}.
\end{equation}

\subsection{BdG spectrum}

$\hat{H}^{2}$ in Eq.~4 in the main text
%\ref{Eq:Squared_Hamilt}

\noindent \begin{align}
%\label{Eq:Squared_Hamilt}
\hat{H}^2= & \sum_{\pmb{k}} \left[ \xi_{+}(\pmb{k}) \tau_{0} + \left( \vec{B}_{\pmb{k}}
\cdot \vec{\tau} \right) \right]^2 \otimes \gamma_{0} + \left| \vec{d}(\pmb{k}) \right|^2 \tau_{0} \otimes \gamma_{0} \notag \\
 & + 2i \left( \vec{B}(\pmb{k}) \times \vec{d}(\pmb{k}) \right) \cdot \vec{\tau} \otimes i \gamma_{2}.
\end{align}

\noindent can be brought to a block-diagonal form in the Nambu indices by applying the transformation

\noindent \begin{equation}
\tilde{U}=e^{-i \gamma_{1} (\pi/4)} \otimes \tau_{0} \times \sigma_{0}
\end{equation}

\noindent such that

\noindent \begin{align}
\label{Eq: H_sqrd_explct}
\tilde{U} \left( \hat{H}^{2} \right) \tilde{U}^{\dagger}=
\begin{pmatrix}
\tilde{H} & 0 \\
0 & \tilde{H}^{T}
\end{pmatrix}
\end{align}

\noindent where

\noindent \begin{align}
\tilde{H}=
\begin{pmatrix}
\xi^{2}_{+}+ | \vec{B} |^{2} + \left|d \right|^{2} + 2 B_{3} \xi_{+} & 2 B_{1}( \xi_{+} - i d_{3}) \\
2 B_{1} ( \xi_{+} + i d_{3}) &  \xi^{2}_{+}+ | \vec{B} |^{2} + \left|d \right|^{2} - 2 B_{3} \xi_{+} .
\end{pmatrix}
\end{align}

%\begin{widetext}

\noindent From this expression,
one can easily check that the eigenvalues of $\hat{H}$
are given by

%\noindent \begin{widetext}
\noindent
\begin{equation}
\label{Eq:Gnrl_2_orbtl_dsprsn}
E_{\pm}(\pmb{k})= \sqrt{ \xi^{2}_{+}(\pmb{k}) + \left| \vec{B}(\pmb{k}) \right|^2
+  \left| \vec{d}(\pmb{k}) \right|^2  \pm \sqrt{ 4\xi^{2}_{+}(\pmb{k})\left| \vec{B}(\pmb{k}) \right|^2 + 4\left| \vec{B}(\pmb{k})
\times \vec{d}(\pmb{k}) \right|^2  }.  }
\end{equation}

\noindent
The explicitly positive semi-definite form of Eq.~5 in the main text
%~\ref{Eq:Balian_Wert_dispersion}
 was obtained by writing

\noindent \begin{align}
\label{Eq:Expndd_sqrt_rt}
\sqrt{ 4\xi^{2}_{+}(\pmb{k})\left| \vec{B}(\pmb{k}) \right|^2 + 4\left| \vec{B}(\pmb{k}) \times \vec{d}(\pmb{k}) \right|^2  }
= & \sqrt{ 4\xi^{2}_{+}(\pmb{k})\left| \vec{B}(\pmb{k}) \right|^2 + 4 B^{2}_{1}(\pmb{k}) d^{2}_{3}(\pmb{k})  } \notag \\
= & 2 \left| \vec{B}(\pmb{k}) \right| \sqrt{ \xi^{2}_{+}(\pmb{k})
+ \frac{B^{2}_{1}(\pmb{k})}{\left| \vec{B}(\pmb{k}) \right|^{2}} d^{2}_{3}(\pmb{k})  } \notag \\
= & 2 \left| \vec{B}(\pmb{k}) \right| \sqrt{ \xi^{2}_{+}(\pmb{k}) + \sin^{2}\phi(\pmb{k}) \left| \vec{d}(\pmb{k}) \right|^{2}  }.
\end{align}
% \end{widetext}
%\end{widetext}

The square can be completed by adding and subtracting
$\sin^{2}\phi(\pmb{k}) \left| \vec{d}(\pmb{k}) \right|^{2}$.

Alternately, a more conventional form for the BdG dispersion
can be obtained from Eq.~\ref{Eq:Gnrl_2_orbtl_dsprsn}
by adding and subtracting $2  \xi_{+}(\pmb{k})  \left| \vec{B}(\pmb{k}) \right|$ to Eq.~\ref{Eq:Expndd_sqrt_rt},
and completing the square for the non-interacting bands $\epsilon^{2}_{\pm}$.
The result is:

\noindent \begin{equation}
\label{Eq:Effct_prng_gps}
E_{\pm}(\pmb{k})= \sqrt{\epsilon^{2}_{\pm}(\pmb{k}) + |\vec{d}|^{2}(\pmb{k}) \pm |Q|(\pmb{k})},
\end{equation}

\noindent where

\noindent \begin{align}
\label{Eq:Fr_dsprsn}
\epsilon_{\pm}(\pmb{k})= & \xi_{+}(\pmb{k}) \pm |\vec{B}|(\pmb{k}) \notag \\
= &  \xi_{+}(\pmb{k}) \pm \sqrt{\xi_{-}^{2}(\pmb{k}) + \xi^{2}_{xy}(\pmb{k})},
\end{align}

\noindent are the electron bands, and the additional $|Q|$ factor is given by

\noindent \begin{equation}
\label{Eq:Efftv_gp}
|Q|(\pmb{k})=  2|\vec{B}|(\pmb{k}) \left( \sqrt{\xi^{2}_{+}(\pmb{k}) + \left| \vec{d}(\pmb{k}) \right|^2 \text{sin}^2\phi(\pmb{k}) }
- \xi_{+}(\pmb{k}) \right).
\end{equation}

\noindent
The presence of this additional
contribution, due to the non-commuting aspect discussed
in the main text,
induces a splitting between the two
conventionally-gapped BdG bands.

\noindent
Indeed, if $\left[\hat{H}_{\text{Kinetic}}, \hat{H}_{\text{Pair}} \right] \sim \vec{B} \times \vec{d}$
(Eq.~4 in the main text)
%~\ref{Eq:Squared_Hamilt})
were to vanish for all $\pmb{k} \in BZ$,
the splitting given by $|Q|$  term would be absent as well.
This can occur for a $\vec{B}$ vector which is either
identically zero or aligned parallel/anti-parallel to
$\vec{d}$ for all momenta. In such cases,
the remaining first two terms in Eq.~\ref{Eq:Effct_prng_gps}
would correspond to a quasiparticle spectrum
with gaps determined by the amplitude of the pairing,
or by the square of the $g_{x^{2}y^{2}}$ form factor in our case.
The resulting BdG bands would be identical to
those for a simpler $s_{x^{2}y^{2}} \otimes \tau_{0}$ state,
which is an example of the $s\pm$ pairing.
As in this latter case, nodes would appear only
when the form factor vanishes along the
$\{\pm \pi/2, k_{y} \}, \{k_{x}, \pm \pi/2 \}$ lines.
A FS which does not intersect these lines would consequently be
completely gapped.
The presence of the last term in Eq.~4 in the main text
%\ref{Eq:Squared_Hamilt}
modifies this simple picture, by introducing
the additional splitting of the two conventionally-gapped BdG
bands.
Furthermore, it is possible that this splitting can be sufficiently strong
to induce nodes for the $\epsilon_{-}$ band. As shown by
Eq.~5 in the main text,
%\ref{Eq:Balian_Wert_dispersion},
these can emerge
along the diagonals $|k_{x}|=|k_{y}|$ of the BZ.
However, we stress that, along the FS, this cannot occur, as explained above.
We also briefly mention that terms similar to $|\vec{Q}|$ are also known in the context
of non-unitary, spin-triplet, time-reversal-symmetry breaking pairings~\cite{S_Sigrist_Ueda:1991}.

\subsection{Band basis}

The pairing Hamiltonian ($\hat{H}_{\text{pair}}$) in the band-basis (Eq.~8 in the main text)
%~\ref{Eq:Pairing_band})
 was obtained from

\noindent \begin{equation}
\left( V(\pmb{k}) \otimes \sigma_{\sigma_{0}}\right) \hat{H}_{\text{pair}}(\pmb{k}) \left( V^{T}(\pmb{k}) \otimes \sigma_{0}\right),
\end{equation}

\noindent where

\noindent \begin{equation}
V(\pmb{k})=
\begin{pmatrix}
\frac{\xi_{-}-\sqrt{\xi_{-}^2+\xi_{xy}^2}}{\sqrt{ \xi^2_{xy}+ \left( \xi_{-} - \sqrt{ \xi_{- }^2+ \xi_{xy}^2 } \right)^2 }} & \frac{\xi_{-}+ \sqrt{\xi_{-}^2+\xi_{xy}^2}}{\sqrt{ \xi^2_{xy}+ \left( \xi_{-} + \sqrt{ \xi_{- }^2+ \xi_{xy}^2 } \right)^2 }}\\
\frac{\xi_{xy}}{\sqrt{ \xi^2_{xy}+ \left( \xi_{-} - \sqrt{ \xi_{- }^2+ \xi_{xy}^2 } \right)^2 }} & \frac{\xi_{xy}}{\sqrt{ \xi^2_{xy}+ \left( \xi_{-} + \sqrt{ \xi_{- }^2+ \xi_{xy}^2 } \right)^2 }}.
\end{pmatrix}.
\end{equation}

\noindent is chosen such that $ V \hat{H}_{\text{Kinetic}} V^{\dagger}$ is diagonal. It can be recast as

\noindent \begin{equation}
\label{}
V(\pmb{k})= \frac{1}{\sqrt{2}}
\begin{pmatrix}
- \sqrt{1-\cos{\phi(\pmb{k})}} &  \sqrt{1+\cos{\phi(\pmb{k})}}\\
\sqrt{1+\cos{\phi(\pmb{k})}} & \sqrt{1- \cos{\phi(\pmb{k})}}
\end{pmatrix},
\end{equation}

\noindent where

\noindent \begin{equation}
\cos \phi(\pmb{k})= \frac{\xi_{-}(\pmb{k})}{\sqrt{\xi_{-}^{2}+ \xi_{xy}^{2}}}.
\end{equation}

\noindent The transformation on $\hat{H}_{\text{Pair}}$ is formally equivalent to the improper rotation

\noindent \begin{equation}
\label{}
\vec{d}'(\pmb{k})=
\begin{pmatrix}
-\cos(\theta) & 0 & \sin(\theta) \\
0 & -1 & 0 \\
\sin(\theta) &  0 & \cos{\theta}
\end{pmatrix}
\begin{pmatrix}
0 \\
0 \\
d
\end{pmatrix}
\end{equation}

\noindent of $\vec{d}(\pmb{k})$ provided that $\theta(\pmb{k})= \phi(\pmb{k})+ \pi$.

\section{The five-orbital model and its solution}
\label{SI:5_orbtl}

\subsection{Model}

We proceed to describe the effective $t-J_{1}-J_{2}$ model we used in our calculations.
These were done for an effective 1-Fe unit cell or equivalently in an unfolded BZ~\cite{S_Nica_Yu_Si:Unpublished_2015}.
To simplify our analysis,
 we consider the kinetic part for all $d$ orbitals but restrict the exchange couplings
 and hence the pairing interactions to $d_{xz}, d_{yz}$, and $d_{xy}$ orbitals only.
 Specifically, the Hamiltonian in the orbital basis is given by

%\noindent \begin{widetext}
\begin{align}
\label{Eq:t_J_Hamiltonian}
H=  -  & \sum_{i<j} (t_{ij}^{\alpha\beta}c_{\alpha}^{\dagger}c_{\beta} + H.C.) + \sum_{i,\alpha} \left(  \epsilon_{i\alpha} - \mu \right) n_{i}+  \sum_{<ij>,\alpha, \beta}J_{1}^{\alpha\beta} \left( \pmb{S}_{i\alpha} \cdot \pmb{S_{j\beta}} - \frac{1}{4}n_{i\alpha}n_{j\beta} \right) +  \notag \\
& + \sum_{<<ij>>,\alpha, \beta}J_{2}^{\alpha\beta} \left( \pmb{S}_{i\alpha} \cdot \pmb{S_{j\beta}} - \frac{1}{4}n_{i\alpha}n_{j\beta} \right)
\end{align}

\noindent \begin{equation}
\label{Eq:Orbital_aniso}
J^{xz/yz}_{1,2} \neq  J^{xy}_{1,2}
\end{equation}
%\end{widetext}

\noindent where $\alpha,\beta \in \{1,2,3,4,5 \}$ are orbital indices representing all five $d_{xz}$, $d_{yz}$, $d_{x^2-y^2}$, $d_{xy}$, and $d_{3z^2-r^2}$ orbitals, $\epsilon_{i}$ are the on-site energies, and $\mu$ is the chemical potential. The local moments can be written as $\pmb{S}_{i \alpha}=\sum_{ss'}\frac{1}{2}c_{i \alpha s}^{\dagger}\pmb{\sigma}_{ss'}c_{i \alpha s'}$ in terms of the conduction electrons. We first consider only intra-orbital exchange ($\alpha=\beta$) and set $J^{x^2-y^2}_{1(2)}=J^{3z^2-r^2}_{1(2)}=0$. We consider general exchange couplings which reflect the possible orbital selectivity by allowing  $J_{xz,xz}=J_{yz,yz} \neq J_{xy,xy}$ (Eq. \ref{Eq:Orbital_aniso}).
The density of states projected onto the $3d_{xy}$ orbital is considerably narrower than that projected onto the $3d_{xz}/3d_{yz}$ orbitals
(with a ratio of about $0.6$ for the alkaline iron selenides)
~\cite{S_Yu_Zhu_Si:PRL_2013}.
Using the square of this ratio as a rough guide, we can expect $A_O
=J_{1}^{xy}/J_{1}^{xz/yz}=J_{2}^{ xy}/J_{2}^{xz/yz}$ to be significantly smaller than $1$ in the iron selenides.

\subsection{Solution method and superconducting pairing phase diagram}

The interactions in Eq. \ref{Eq:t_J_Hamiltonian} can be decomposed into nearest-neighbor (NN) and next-nearest neighbor (NNN) singlet pairing terms. The double occupancy constraint can be incorporated in practice through a band renormalization by the doping factor $\delta= \left|\sum_{i,s} n_{i\alpha s}- 2 \right|$. The pairing Hamiltonian can be solved numerically in a 1-Fe unit cell calculation by varying the exchange couplings.
For more details on the method, we refer the reader to Refs. \onlinecite{Yu_Nat_Comm:2013,Yu_Zhu_Si:2014}.
As specified above,
an exchange orbital anisotropy factor
is defined as $A_{O}=J_{1}^{xy}/J_{1}^{xz/yz}=J_{2}^{ xy}/J_{2}^{xz/yz}$
and an orbital-independent NN-NNN exchange anisotropy factor $A_{L}=J_{1}^{\alpha}/J_{2}^{\alpha}$
for all three non-zero intra-orbital exchange couplings for $d_{xz}$, $d_{yz}$, and $d_{xy}$.

\begin{figure}[!h]
\centering
\includegraphics[width=0.4\columnwidth]{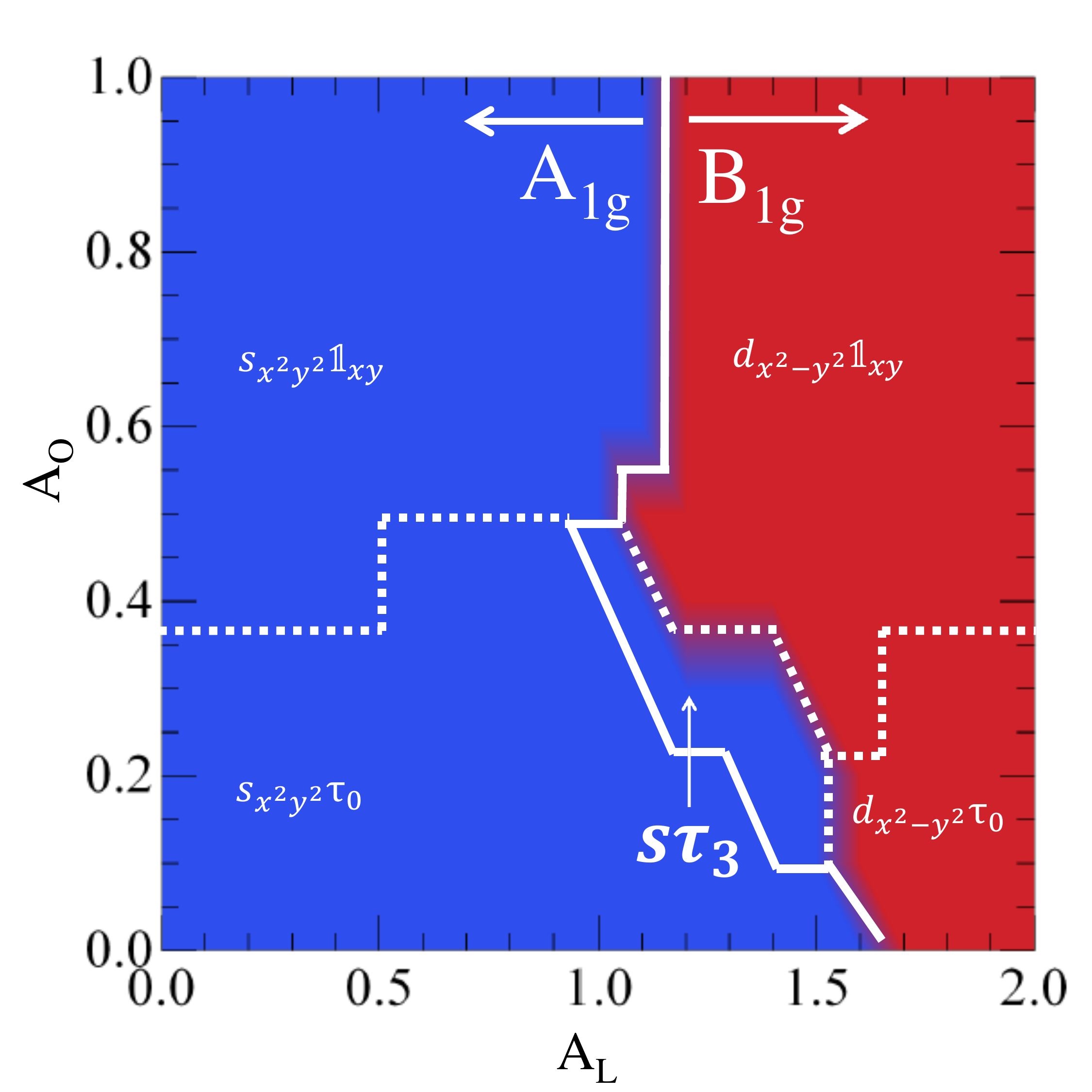}
\caption{ Phase diagram based on the leading pairing amplitudes given by self-consistent calculations using tight-binding parameters appropriate to single-layer FeSe. The tight-binding parameters used can be found in
Ref.~\onlinecite{Yu_Nat_Comm:2013}. The blue shaded areas correspond to dominant pairing channels with an $s_{x^2y^2}$ form factor while the red shading covers those with a $d_{x^2-y^2}$ form factor. The continuous line separates regions where the pairing belongs to the $A_{1g}$ and the $B_{1g}$ representations respectively. The $1 \times 1$ matrix in the $d_{xy}$ subspace is represented by $\pmb{1}_{xy}$.}
\label{Phase_diagram_single_layer_FeSe_PNAS_11_10_2016}
\end{figure}

\begin{figure}[!h]
\centering
\includegraphics[width=0.4\columnwidth]{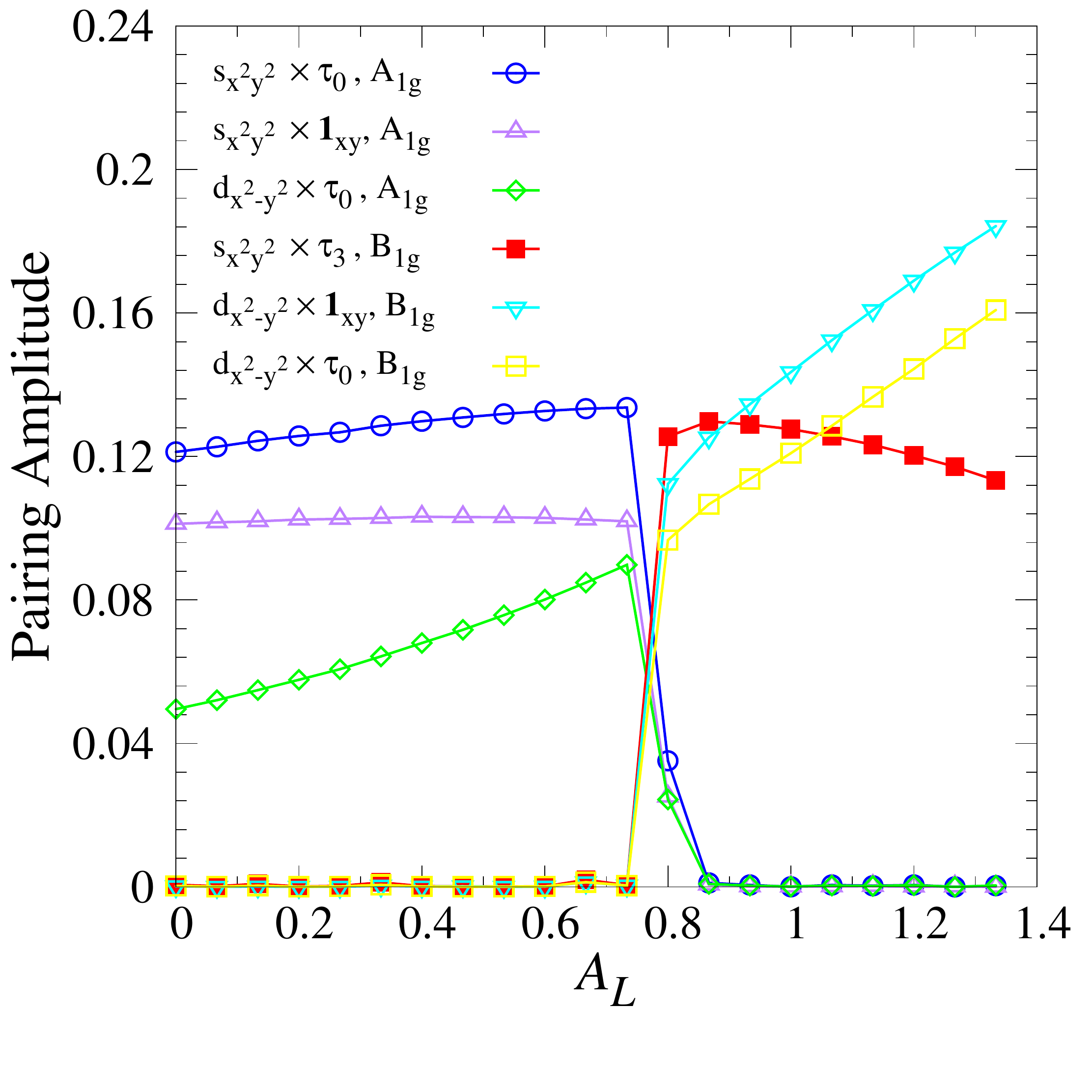}
\caption{Leading pairing amplitudes (vertical axis) for a dispersion typical of alkaline iron selenides for fixed  $J_2=1.5, A_O=0.3$ and varying NN-NNN ratio $A_L$ (horizontal axis). The $\tau$ label indicates a dominant $d_{xz}, d_{yz}$ sector while $ \mathbf{1}_{xy}$ marks a $d_{xy}$ dominant pairing. For $ 0.8 \leq A_L \leq 0.94 $ the leading pairing is in the $s\tau_{3}$ channel shown in dark filled squares. Note that the reduced parameter space for the $s\tau_{3}$ is due to the proximity to the phase boundary and for lower values of $A_O$ the range over which this pairing leads is increased.}
\label{Fig:Pairing_amplitude_A_O=0.5_A_L=1.3}
\end{figure}

To explore the zero-temperature superconducting phases corresponding to different classes of Fe-based materials
we consider the associated electron dispersions for
$ \textnormal{K}_{\textnormal{y}}\textnormal{Fe}_{\textnormal{2-x}}\textnormal{Se}_{\textnormal{2}} $, iron pnictides
and single-layer FeSe. We subsequently tune the exchange couplings for various NN-NNN and orbital anisotropy ratios
($A_L$ and $A_O$) and determine the real-space pairing functions.
This leads to the pairing phase diagram in the $A_L-A_O$ parameter space. The results for the electronic dispersions of the alkaline iron selenides and iron pnictides are shown in the main text as Figs.~2 (a) and (b), respectively. Those for the case of the single-layer FeSe is shown here, in Fig.~\ref{Phase_diagram_single_layer_FeSe_PNAS_11_10_2016}. For the case of the alkaline iron selenides, a cut along the $A_L$ axis for a fixed $A_O=0.3$ is shown in Fig.~\ref{Fig:Pairing_amplitude_A_O=0.5_A_L=1.3}.

\subsection{Effects of inter-orbital exchange interactions}

Throughout the main text, the discussion has been centered on cases with only
intra-orbital
$J$'s and their consequence on the pairing amplitudes.
To analyze the robustness of our results, we turn to
calculations which allow for
\emph{inter-orbital} NN and NNN ($J_{1}$ and $J_{2}$, respectively) exchange interactions
between the dominant $d_{xz}, d_{yz}$, and $d_{xy}$ orbitals, in addition to the intra-orbital interactions
considered in Eq.~\ref{Eq:t_J_Hamiltonian}. More specifically, we introduce

\noindent \begin{align}
J^{xz/yz}_{1} = & J^{yz/xz}_{1} = 0 \notag \\
J^{xz/xy}_{1} = & \sqrt{ J^{xz/xz}_{1} \times J^{xy/xy}_{1} } \notag \\
J^{yz/xy}_{1} = & J^{xz/xy}_{1}
\end{align}

\noindent and

\noindent

\begin{align}
J^{xz/yz}_{2} = & J^{xz/xz}_{2} =  J^{yz/yz}_{2} \notag \\
J^{xz/xy}_{2} = & \sqrt{ J^{xz/xz}_{2} \times J^{xy/xy}_{2} } \notag \\
J^{yz/xy}_{2} = & J^{xz/xy}_{2}
\end{align}

Crucially, these conditions allow the inter-orbital coupling constants to be consistent with the underlying
super-exchange
mechanism. Thus, the absence of NN hopping between $d_{xz}$ and $d_{yz}$ orbitals~\cite{S_Yu_Nat_Comm:2013,Yu_Zhu_Si:2014} implies vanishing $J^{xz/yz}_{1}, J^{yz/xz}_{1}$.
Similarly, the $xz-xy$ and $yz-xy$ super-exchange coupling constants, involving the square-root terms, reflect the influence of orbitally-selective correlations.

\begin{figure}[h!]
\centering
\subfloat[]{\includegraphics[width=0.4\columnwidth]{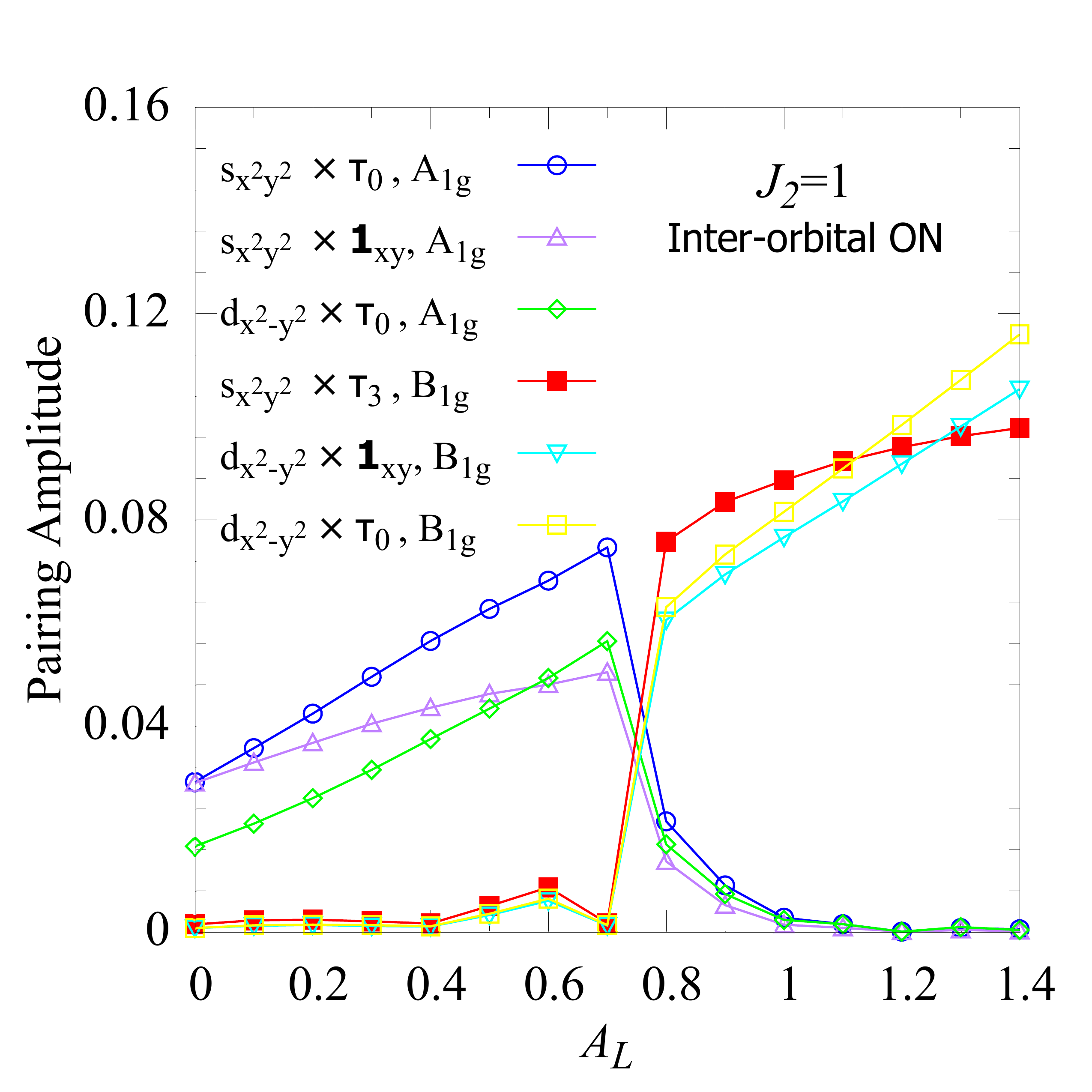}} \\
\subfloat[]{\includegraphics[width=0.4\columnwidth]{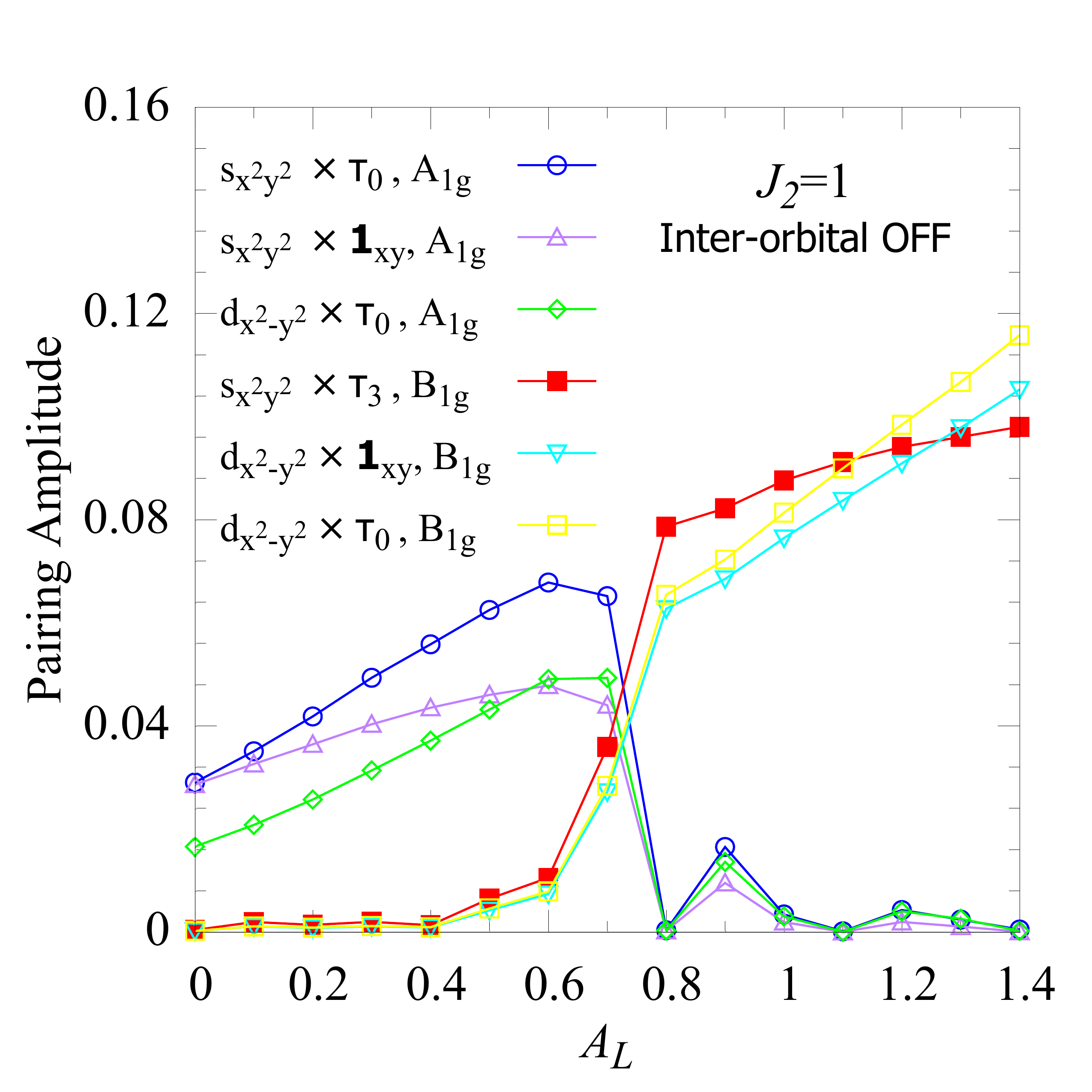}}
\caption{Leading intra-orbital pairing amplitudes (vertical axis) for a dispersion typical of alkaline iron selenides for fixed  $J_2=1, A_O=0.2$ and varying NN-NNN ratio $A_L$ (horizontal axis) \emph{with (a) and without (b) inter-orbital exchange interactions}. As mentioned in the discussion above, no significant changes are observed.
}
\label{Fig:K_Prng_ampltds_intra}
\end{figure}

In Figs.~\ref{Fig:K_Prng_ampltds_intra} (a) and (b), we show the amplitudes for the leading \emph{intra-orbital} pairing channels in the case of the alkaline iron selenides, for $A_{O}=0.2$ and $J_{2}=1$, with and without inter-orbital exchange interactions. As these figures clearly show, no significant changes occur. Similar pictures emerge for virtually all values of $A_{O}$ and $A_{L}$ shown in the phase diagram in Fig.~2 (a) in the main text.

\begin{figure}[h!]
\centering
\subfloat[]{\includegraphics[width=0.4\columnwidth]{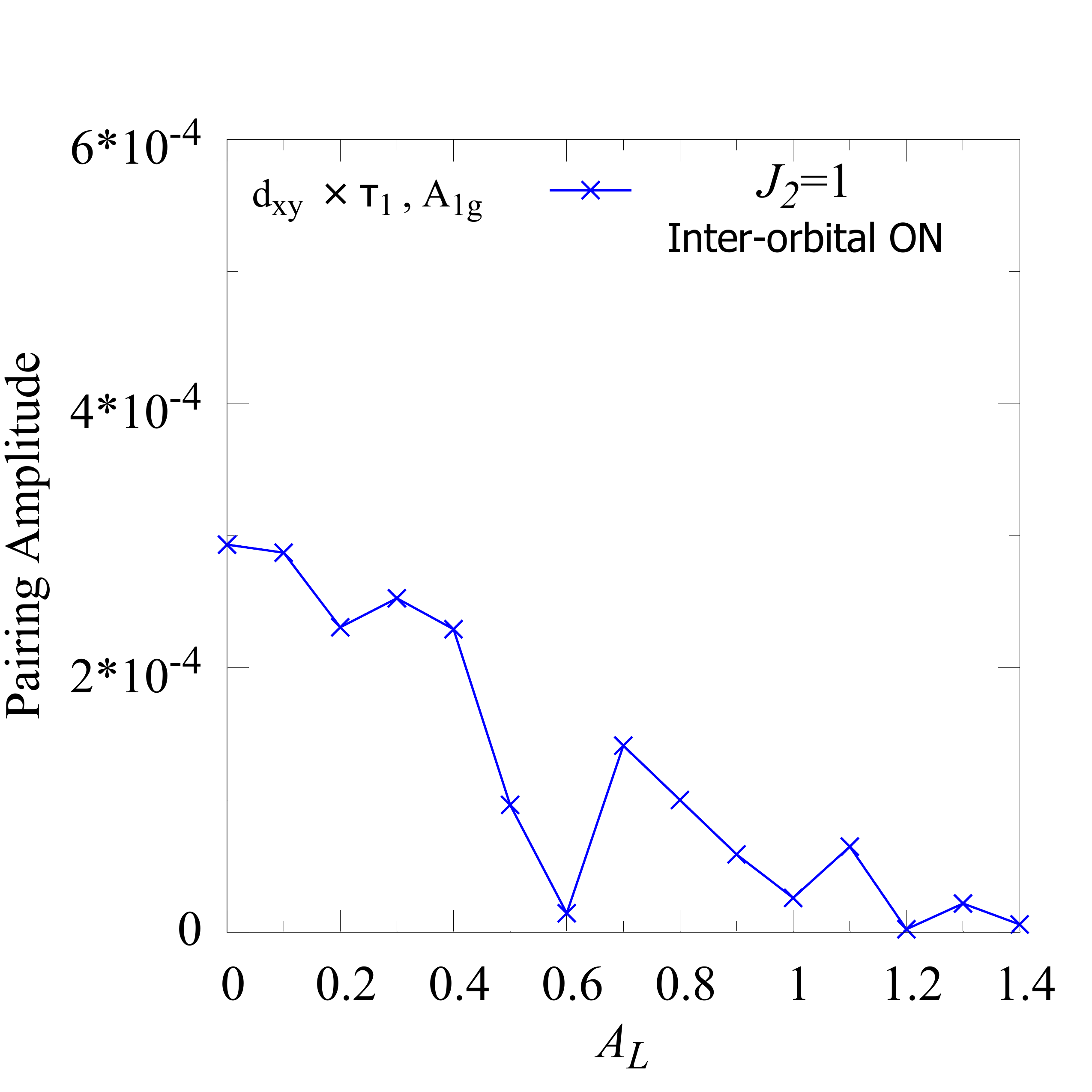}} \\
\subfloat[]{\includegraphics[width=0.4\columnwidth]{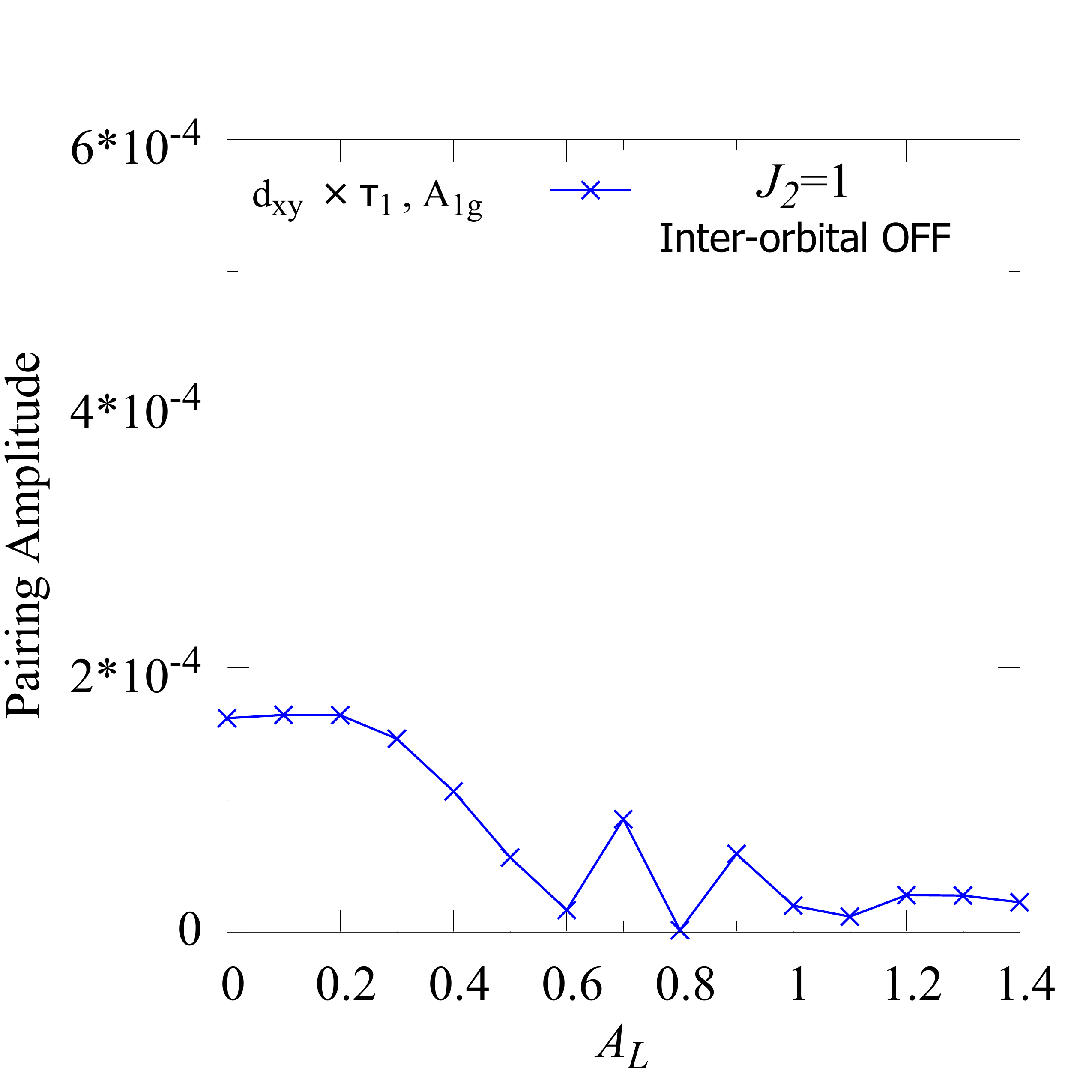}}
\caption{Leading inter-orbital pairing amplitude (vertical axis) for a dispersion typical of alkaline iron selenides for fixed  $J_2=1, A_O=0.2$ and varying NN-NNN ratio $A_L$ (horizontal axis) \emph{with (a) and without (b) inter-orbital exchange interactions}. As mentioned in the discussion above, no significant changes are observed.
}
\label{Fig:K_Prng_ampltds_inter}
\end{figure}

In Figs.~\ref{Fig:K_Prng_ampltds_inter} (a) and (b), we plot
one of the leading \emph{inter-orbital}
pairing amplitudes for the alkaline iron selenides,
in the $d_{xy} \otimes \tau_{1}, A_{1g}$ channel,
with and without inter-orbital exchange couplings. In either case, the leading inter-orbital pairing amplitude is roughly two orders of magnitude smaller than the leading intra-orbital amplitude.
(The numerical accuracy of our calculation for the pairing amplitudes is about $10^{-4}$.)
The same conclusion is drawn throughout the phase diagram.

Based on these results and similar ones for the Fe-pnictide cases, we conclude that
the
inter-orbital exchange interactions
have
a negligible effect on the pairing amplitudes within our model.

\section{Dynamical spin susceptibility and neutron resonance}
\label{Sec:Grnl_rqrmnt_spn_rsnnc}

\subsection{General formulation}
In the single-band BCS case, the
bare
contribution to the dynamical spin susceptibility (see Eq.~\ref{Eq:RPA_suscept} for the multi-orbital case) depends~\cite{S_Eschrig:Adv_Phys_2006, S_Fong_et_al:PRL_1995} on terms like

%\noindent \begin{widetext}
\begin{align}
\label{Eq:BCS_RPA_suscept}
\chi_{0}(\pmb{q},\omega)=\frac{1}{N}\sum_{\pmb{k}} \bigg[ \frac{1}{2} \bigg( 1- \frac{\epsilon_{\pmb{k}+\pmb{q}} \epsilon_{\pmb{k}} + \Delta_{\pmb{k}+\pmb{q}} \Delta_{\pmb{k}} }{E_{ \pmb{k} + \pmb{q} } E_{\pmb{k}} } \bigg) \frac{ f( E_{ \pmb{k} + \pmb{q} }) + f(E_{\pmb{k}}) -1 }{ \omega - ( E_{ \pmb{k} + \pmb{q} }+ E_{ \pmb{k} } ) + i0^{+} } + ... \bigg] ,
\end{align}
%\end{widetext}

\noindent where $\epsilon$'s and $E$'s are the free particle and the BdG quasi-particle dispersions respectively.
The existence of a sharp feature in the RPA
dynamical spin susceptibility
below the particle-hole threshold
(given roughly by twice the characteristic gap magnitude $2\Delta$) is related to the sign
of the $\Delta_{\pmb{k}+\pmb{q}} \Delta_{\pmb{k}}$ term
in the spin (time-reversal-odd) coherence factor in Eq.~\ref{Eq:BCS_RPA_suscept}.
Close to the Fermi surface, when the sign is positive,
the coherence factor suppresses
$\chi_{0}(\pmb{q},\omega)$ and, consequently,
inhibits the appearance of a resonance.
By contrast, when $\Delta_{\pmb{k}+\pmb{q}}$ and $\Delta_{\pmb{k}}$ have opposite signs, the resonance can form
at an energy below $2\Delta$.

In the present multi-orbital model,
the bare dynamical spin susceptibility is defined as

\noindent \begin{equation}
\label{Eq:RPA_suscept}
\chi_0(\pmb{q}, i\omega_n)=\sum_{\alpha \beta} \chi_{0 \alpha \beta}(\pmb{q}, i\omega_n),
\end{equation}

\noindent where

\noindent \begin{equation}
\chi_{0, \alpha \beta}(\pmb{q}, i\omega_n)=\int_{1}^{1/T} d\tau e^{i \omega_{n} \tau } \left< \mathcal{T}_{\tau}
\left[ S^{-}_{ \pmb{q}\alpha}(\tau)S^{+}_{ \pmb{-q}\beta}(0) \right] \right>.
\nonumber
\end{equation}

\noindent The interaction corrected susceptibility is then
\noindent \begin{align}
\chi_{\alpha \beta}(\pmb{q}, i\omega_n)=\sum_{\gamma} & \left[ \textbf{I} + J(\pmb{q}) \sum_{\delta \mu}
\chi_{0, \delta \mu} (\pmb{q}, i\omega_n) \right]^{-1}_{\alpha \gamma} \times \notag \\
& \times \chi_{0, \gamma \beta} (\pmb{q}, i\omega_n),
\nonumber
\end{align}

\noindent where

\noindent \begin{equation}
J(\pmb{q})=\frac{J_1}{2} \left( \text{cos} q_x + \text{cos} q_y \right) + J_2 \text{cos} q_x \text{cos} q_y.
\nonumber
\end{equation}

In our case, the intraband pairing component has a sign change across the electron Fermi pockets at the BZ edges.
This implies that the corresponding component of the bare susceptibility $\chi_0$ will dominate the final contribution to
the imaginary part of the renormalized spin susceptibility, ${\rm Im}\chi$, at the wavevector $(\pi,\pi/2)$,
which spans across the two electron Fermi pockets. We
discuss this issue further in the next subsection.

\subsection{Spin resonance in the alkaline iron selenides}
Here
the dynamical spin susceptibility of interest is near the wave vector $\pmb{q}$
which connects the two electron pockets near the BZ boundaries $(\pi,0)$ and $(0,\pi)$  [Fig.~4a, main text].
While both the intraband and interband components of the pairing function are crucial for the overall properties of the $s \tau_3$ pairing state,
as far as the spin resonance is concerned, the involved electron Fermi pockets
near $(\pi,0)$ and $(0,\pi)$
 belong to only one band.
We can then treat the ratio of the interband pairing amplitude to the separation of the energies between the neighboring (normal state)
energy bands as a perturbation. In this way, we obtain a simplified expression for the leading term of the dynamical susceptibility,
which links the spin resonance with the sign change of the intraband
 component of the pairing function.

In the band basis, the bare dynamical spin susceptibility is written as
\noindent \begin{equation}
\label{Eq:chi_0_multiband}
\chi_{0}(\pmb{q}, i\omega_n)=\frac{1}{N} \sum_{\pmb{k}} \sum_{a,b} \mathfrak{F}_{\pmb{k},\pmb{q},a,b} \frac{f(E_{\pmb{k},a})+f(E_{\pmb{k+q},b})-1}{i\omega_n-E_{\pmb{k},a}-E_{\pmb{k}+\pmb{q},b}},
\end{equation}
\noindent where $a$ and $b$ run over all the
BdG quasiparticle bands, and $\mathfrak{F}_{\pmb{k},\pmb{q},a,b}$ is a
prefactor
with the following generic expression,

%\noindent \begin{widetext}
\noindent \begin{equation}
\label{Eq:Coherence_Factor}
 \mathfrak{F}_{\pmb{k},\pmb{q},a,b} = \left[ \sum_{B,D}\tilde{V}_{BD,\pmb{k},\pmb{q}} \bar{U}^\star_{\pmb{k}(B\downarrow,a)} \bar{U}^\star_{\pmb{k+q}(D\uparrow,b)} \right]
 \left[ \sum_{A,C} \tilde{V}_{AC,\pmb{k},\pmb{q}} \bar{U}_{\pmb{k}(A\downarrow,a)} \bar{U}_{\pmb{k+q}(C\uparrow,b)} - \tilde{V}^\star_{AC,\pmb{k},\pmb{q}} \bar{U}_{\pmb{k}(A\uparrow,a)} \bar{U}_{\pmb{k+q}(C\downarrow,b)} \right] .
\end{equation}
%\end{widetext}

\noindent Here $A$-$D$ are the indices of the bands in the normal state.
$\tilde{V}_{AC,\pmb{k},\pmb{q}}=\sum_{\alpha} V_{\alpha A}(\pmb{k}) V^\star_{\alpha C}(\pmb{k+q})$ is a
factor
associated with the canonical transformation $V(\pmb{k})$ from the orbital basis to the band basis.
This factor describes the band-dependent contribution to the spin operator and
have the same form in the normal and superconducting states.
$\bar{U}_{\pmb{k}(A\sigma,a)}$ is a matrix element of the Bogoliubov transformation that diagonalizes the pairing Hamiltonian in the band basis (see below).

We denote by ``$+$"
 the (normal state) band that crosses the Fermi energy near $(\pi,0)$ and $(0,\pi)$ [{\it cf.} Fig.~4a, main text].
In general, the interband pairing amplitude will be small compared to the separations of this band from the other bands near that part of the 
%\ry{1-Fe} BZ. 
1-Fe BZ.
We can then simplify the analysis
by considering the $+$ band along with only a second band, denoted by ``$-$".
The effective Hamiltonian reads
\begin{equation}
 \hat{H}=
 \begin{pmatrix}
  \epsilon_+ & 0 & \Delta_{++} & \Delta_{+-} \\
  0 & \epsilon_- & \Delta_{+-} & \Delta_{--} \\
  \Delta_{++} & \Delta_{+-} & -\epsilon_+ & 0 \\
  \Delta_{+-} & \Delta_{--} & 0 & -\epsilon_-
 \end{pmatrix},
\end{equation}

\noindent where $\epsilon_{+/-}$ are the energies of the bands in the normal state;
 $\Delta_{++}$ and $\Delta_{--}$ are the intraband pairing components;
  $\Delta_{+-}$ is the interband pairing component,
  satisfying the condition $ | \Delta_{+-} | \ll |\epsilon_{+}-\epsilon_{-}|$.
  The Hamiltonian can be diagonalized by a Bogoliubov transformation $\bar{U}_{\pmb{k}}$,
  and we obtain the excitation energies of the
  BdG energy dispersion
%\noindent \begin{widetext}
\noindent \begin{equation}
E_{\pm}(\pmb{k})=\sqrt{\frac{1}{2} \left\{ \epsilon_+^2+\epsilon_-^2+\Delta_{++}^2+\Delta_{--}^2+2\Delta_{+-}^2 \pm \sqrt{\left[ \epsilon_+^2-\epsilon_-^2+\Delta_{++}^2-\Delta_{--}^2 \right]^2 + 4\Delta_{+-}^2 \left[ (\epsilon_+-\epsilon_-)^2 +(\Delta_{++}-\Delta_{--})^2  \right]} \right\}}.
\end{equation}
%\end{widetext}
\noindent We stress again that
we are focusing on the pairing near the electron pockets centered at $(\pi,0)$ and $(0,\pi)$ only.
Because, in the 1-Fe BZ, only one band crosses the Fermi level at these electron Fermi pockets~\cite{S_Yu_Nat_Comm:2013} and the energy separation between this band and nearby hole band is about $100$ meV,\cite{S_MYi.2015} which is much larger than the pairing functions,
there is a strong constraint to the summations in Eqs.~\ref{Eq:chi_0_multiband} and \ref{Eq:Coherence_Factor}: In Eq.~\ref{Eq:chi_0_multiband},
the relevant term
of the dynamical spin susceptibility
 is now the one with $a=b=+$, and in Eq.~\ref{Eq:Coherence_Factor}
  the term with $A=B=C=D=+$ contributes the most to the
  prefactor because the energy separation of the bands $|\epsilon_+-\epsilon_-|$ is much larger than the pairing components.
As a result, the leading term of the bare dynamical spin susceptibility
 reads

%\noindent \begin{widetext}
\noindent \begin{equation}
\chi_{0}(\pmb{q}, i\omega_n) \sim \frac{1}{N} \sum_{\pmb{k}}
\tilde{V}^2_{++,\pmb{k},\pmb{q}} \bar{U}^\star_{\pmb{k}(+\downarrow,+)} \bar{U}^\star_{\pmb{k+q}(+\uparrow,+)} \left[ \bar{U}_{\pmb{k}(+\downarrow,+)} \bar{U}_{\pmb{k+q}(+\uparrow,+)} - \bar{U}_{\pmb{k}(+\uparrow,+)} \bar{U}_{\pmb{k+q}(+\downarrow,+)} \right]
\frac{f(E_+(\pmb{k}))+f(E_+(\pmb{k+q}))-1}{i\omega_n-E_+(\pmb{k})-E_+(\pmb{k}+\pmb{q})}.
\end{equation}
%\end{widetext}

We define the small parameter $\eta \equiv
\Delta_{+-}/\sqrt{\epsilon^2_- - \epsilon^2_+ + \Delta_{--}^2 -\Delta_{++}^2} $,
and expand
the BdG energy dispersion
$E_{\pm}(\pmb{k})$ and the matrix elements of the Bogoliubov transformation $\bar{U}_{\pmb{k}(A\sigma,a)}$ in terms of $\eta$. We obtain,

%\noindent \begin{widetext}
\noindent\begin{align}
& E_+(\pmb{k}) = E_{+0} + \frac{\eta^2\left[ (\epsilon^2_+ + \Delta^2_{++})-(\epsilon^2_- + \Delta^2_{--}) + (\epsilon_+-\epsilon_-)^2 +(\Delta_{++}-\Delta_{--})^2 \right]}{2E_{+0}} + O(\eta^3)\\
& \bar{U}_{\pmb{k}(+\uparrow,+)} = \sqrt{\frac{E_{+0}+\epsilon_+}{2E_{+0}}} + O(\eta^2)\\
& \bar{U}_{\pmb{k}(+\downarrow,+)} = \sqrt{\frac{E_{+0}-\epsilon_+}{2E_{+0}}} + O(\eta^2)\\
& \bar{U}_{\pmb{k}(-\uparrow,+)} = \frac{\eta}{\sqrt{\epsilon^2_- - \epsilon^2_+ + \Delta_{--}^2 -\Delta_{++}^2}}\left[ (\epsilon_- - E_{+0}) \sqrt{\frac{E_{+0}-\epsilon_+}{2E_{+0}}} -\Delta_{--} \sqrt{\frac{E_{+0}+\epsilon_+}{2E_{+0}}} \right]+ O(\eta^2)\\
& \bar{U}_{\pmb{k}(-\downarrow,+)} = \frac{\eta}{\sqrt{\epsilon^2_- - \epsilon^2_+ + \Delta_{--}^2 -\Delta_{++}^2}}\left[ -\Delta_{--} \sqrt{\frac{E_{+0}-\epsilon_+}{2E_{+0}}} -(\epsilon_- + E_{+0}) \sqrt{\frac{E_{+0}+\epsilon_+}{2E_{+0}}} \right]+ O(\eta^2),
\end{align}
%\end{widetext}

\noindent where $E_{+0}(\pmb{k})=\sqrt{\epsilon_+^2(\pmb{k})+\Delta_{++}^2(\pmb{k})}$.

This leads to the following form for the leading term of $\chi_0$:

%\noindent \begin{widetext}
\noindent \begin{equation}
\label{Eq:chi0_final_form}
\chi_{0} (\pmb{q}, i\omega_n) \sim \frac{1}{N} \sum_{\pmb{k}}
\tilde{V}^2_{++,\pmb{k},\pmb{q}} \frac{1}{2} \bigg( 1- \frac{\epsilon_{+,\pmb{k}+\pmb{q}} \epsilon_{+,\pmb{k}} + \Delta_{++,\pmb{k}+\pmb{q}} \Delta_{++,\pmb{k}} }{E_{+0,\pmb{k} + \pmb{q} } E_{+0,\pmb{k}} } \bigg)
\frac{f(E_{+0}(\pmb{k}))+f(E_{+0}(\pmb{k+q}))-1}{i\omega_n-E_{+0}(\pmb{k})-E_{+0}(\pmb{k}+\pmb{q})} + O(\eta).
\end{equation}
Here, the prefactor $\tilde{V}^2_{++,\pmb{k},\pmb{q}}$ is the same as in the normal state;
it simply weighs the contribution of this particular band to the p-h excitation in the spin channel at these wave vectors.

In Eq.~\ref{Eq:chi0_final_form}, the effect of superconductivity appears through the factor in the big brackets,
which is essentially the same as the spin coherence factor
of the 1-band case, given in Eq.~\ref{Eq:BCS_RPA_suscept} (an analytical continuation $i\omega_n\to\omega+i0^+$ is needed to compare the two
equations).

Similar to the usual case
~\cite{S_Eschrig:Adv_Phys_2006},
a sharp resonance appears in the imaginary part of the dynamical spin susceptibility $\chi''(\pmb{q}, \omega)$
when there is a sign change in the intraband pairing components $\Delta_{++}(\pmb{k})$ between the two electron pockets.
This conclusion is consistent with our numerical result for $\chi''(\pmb{q}, \omega)$, shown in Fig.~3 of the main text.

%\end{widetext}

\noindent 

\noindent 

\noindent 
\noindent 

%\nocite{*}
%\bibliography{tau_3_pnas_02_06_2017_revtex}

%
%%merlin.mbs apsrev4-1.bst 2010-07-25 4.21a (PWD, AO, DPC) hacked
%%Control: key (0)
%%Control: author (0) dotless jnrlst
%%Control: editor formatted (1) identically to author
%%Control: production of article title (0) allowed
%%Control: page (1) range
%%Control: year (0) verbatim
%%Control: production of eprint (0) enabled
%